\documentclass[manuscript,screen]{acmart}
\AtBeginDocument{%
  \providecommand\BibTeX{{%
    Bib\TeX}}}

\usepackage{amsmath,amsfonts}
\usepackage{algorithmic}
\usepackage{graphicx}
\usepackage{textcomp}
\usepackage{xcolor}
\def\BibTeX{{\rm B\kern-.05em{\sc i\kern-.025em b}\kern-.08em T\kern-.1667em\lower.7ex\hbox{E}\kern-.125emX}}
\usepackage{array}
\usepackage[Symbol]{upgreek}
\usepackage{booktabs}
\usepackage{longtable}
\usepackage{mathtools}
\usepackage{tikz}
\usetikzlibrary{tikzmark}
\usetikzlibrary{arrows.meta}
\usetikzlibrary{shapes.geometric, positioning}
\usepackage{multirow}
\usepackage{graphicx}
\usepackage{lscape}
\usepackage{float}
\usepackage{xcolor}
\usepackage[table, svgnames]{xcolor} 
\usepackage[dvipsnames]{xcolor}
\usepackage{tabularx,ragged2e,booktabs,caption}
\usepackage{dblfloatfix}
\usepackage{footnote}
\usepackage{makecell}
\usepackage{xcolor}
\usepackage[utf8]{inputenc}
\usepackage[nonumberlist,nogroupskip,nopostdot]{glossaries}

\makeglossaries
\usepackage{forest,array}
\usepackage{tikz}
\usepackage{soul}
\usepackage{pifont}
\usepackage{diagbox}
\usepackage{pgfplots}
\pgfplotsset{width=7cm,compat=1.3}
\definecolor{Gray}{gray}{0.9}
\definecolor{blizzardblue}{rgb}{0.67, 0.9, 0.93}
\usepackage{tcolorbox}

\usepackage{subcaption}
\usepackage{bytefield}
\usepackage{enumitem}

\setcopyright{acmlicensed}
\copyrightyear{2026}
\acmYear{2026}
\acmDOI{XXXXXXX.XXXXXXX}
\acmConference[Conference acronym 'XX]{Make sure to enter the correct
  conference title from your rights confirmation email}{June 03--05,
  2026}{Woodstock, NY}
\acmBooktitle{ACM Transactions on Privacy and Security, 2026}
\acmISBN{978-1-4503-XXXX-X/2018/06}

\begin{document}
\newacronym{vnf}{VNF}{Virtualized Network Function}
\newacronym{5g}{5G}{Fifth Generation}
\newacronym{5g-advanced}{5G-Advanced}{}
\newacronym{6g}{6G}{Sixth Generation}
\newacronym{mmtc}{mMTC}{massive Machine Type Communications}
\newacronym{embb}{eMBB}{enhanced Mobile Broadband}
\newacronym{urllc}{URLLC}{Ultra-Reliable Low Latency Communications}
\newacronym{sdn}{SDN}{Software Defined Networking}
\newacronym{nfv}{NFV}{Network Function Virtualization}
\newacronym{nf}{NF}{Network Function}
\newacronym{dn}{DN}{Data Network}
\newacronym{arpf}{ARPF}{Authentication credential Repository and Processing Function}
\newacronym{gtp}{GTP}{General Packet Radio Service Tunnelling Protocol}
\newacronym{gprs}{GPRS}{General Packet Radio Service}
\newacronym{teid}{TEID}{Tunnel Endpoint Identifier}
\newacronym{ip}{IP}{Internet Protocol}
\newacronym{icmp}{ICMP}{Internet Control Message Protocol}
\newacronym{dns}{DNS}{Domain Name System}
\newacronym{pcap}{PCAP}{Packet Capture}
\newacronym{pps}{PPS}{Packets Per Second}
\newacronym{rto}{RTO}{Retransmission Timeout}
\newacronym{cwnd}{CWND}{Congestion Window}
\newacronym{sba}{SBA}{Service-Based Architecture}

\newacronym{ietf}{IETF}{Internet Engineering Task Force}

\newacronym{supi}{SUPI}{Subscription Permanent Identifier}
\newacronym{capex}{CAPEX}{Capital Expenditure}
\newacronym{opex}{OPEX}{Operational Expenditure}
\newacronym{dos}{DoS}{Denial-of-Service}
\newacronym{ddos}{DDoS}{Distributed DoS}
\newacronym{minlp}{MINLP}{Mixed-Integer Nonlinear Programming}
\newacronym{milp}{MILP}{Mixed-Integer Linear Programming}
\newacronym{ilp}{ILP}{Integer Linear Programming}
\newacronym{upf}{UPF}{User Plane Function}
\newacronym{mec}{MEC}{Multi-access Edge Computing}
\newacronym{qos}{QoS}{Quality of Service}
\newacronym{sfc}{SFC}{Service Function Chain}
\newacronym{pdu}{PDU}{Protocol Data Unit}
\newacronym{ns}{NS}{Network Slicing}
\newacronym{sla}{SLA}{Service Level Agreement}
\newacronym{slo}{SLO}{Service Level Objective}
\newacronym{dps}{DPS}{Data Plane Services}
\newacronym{cps}{CPS}{Control Plane Services}
\newacronym{dp}{DP}{Data Plane}
\newacronym{cp}{CP}{Control Plane}
\newacronym{vm}{VM}{Virtual Machine}
\newacronym{mip}{MIP}{Mixed Integer Programming}
\newacronym{amf}{AMF}{Access and Mobility Function}
\newacronym{smf}{SMF}{Session Management Function}
\newacronym{nrf}{NRF}{Network Repository Function}
\newacronym{scip}{SCIP}{Solving Constraint Integer Programs}
\newacronym{ue}{UE}{User Equipment}

\newacronym{hplmn}{H-PLMN}{Home \gls{plmn}}
\newacronym{plmn}{PLMN}{Public Land Mobile Network}
\newacronym{eap-aka}{EAP-AKA'}{Extensible Authentication Protocol-Authentication and Key Agreement}
\newacronym{aka}{AKA}{Authentication and Key Agreement}
\newacronym{gnb}{gNB}{gNodeB}
\newacronym{3gpp}{3GPP}{3rd Generation Partnership Project}
\newacronym{cn}{CN}{Core Network}
\newacronym{ran}{RAN}{Radio Access Network}
\newacronym{udm}{UDM}{Unified Data Management}
\newacronym{ausf}{AUSF}{Authentication Server Function}

\newacronym{mno}{MNO}{Mobile Network Operator}

\newacronym{udp}{UDP}{User Datagram Protocol}
\newacronym{tcp}{TCP}{Transmission Control Protocol}

\newacronym{rtt}{RTT}{Round Trip Time}

\newacronym{v2x}{V2X}{Vehicle-to-Everything}
\newacronym{dpdk}{DPDK}{Data Plane Development Kit}
\newacronym{gcn}{GCN}{Graph Convolutional Network}
\newacronym{nsga}{NSGA-II}{Non-Dominated Sorting Genetic Algorithm II}

\newacronym{iot}{IoT}{Internet of Things}
\newacronym{b2b2x}{B2B2X}{Business-to-Business-to-Everything}
\newacronym{vpn}{VPN}{Virtual Private Network}

\newacronym{ccs}{CCS}{Conference on Computer and Communications Security}
\newacronym{vlan}{VLAN}{Virtual Local Area Network}
\newacronym{pfcp}{PFCP}{Packet Forwarding Control Protocol}
\newacronym{uav}{UAV}{Unmanned Aerial Vehicle}
\newacronym{xr}{XR}{Extended Reality}



\title{Slice Isolation-as-a-Service: An Isolation Mechanism to Proactively Defend Cross-Slice Attacks in 5G-Advanced and Beyond\\
}

\subtitle{\normalfont Author’s draft for soliciting feedback -- November 16, 2026}

\author{AbdulAziz AbdulGhaffar}
\orcid{0000-0002-3888-3975}
\affiliation{%
  \institution{Department of Systems and Computer Engineering, Carleton University}
  \city{Ottawa}
  \state{Ontario}
  \country{Canada}
}
\email{abdulazizabdulghaff@cmail.carleton.ca}

\author{Ashraf Matrawy}
\orcid{0000-0001-9220-4630}
\affiliation{%
  \institution{School of Information Technology, Carleton University}
  \city{Ottawa}
  \state{Ontario}
  \country{Canada}
}
\email{AshrafMatrawy@cunet.carleton.ca}

\renewcommand{\shortauthors}{AbdulGhaffar et al.}

\begin{abstract}
  As cellular networks transition toward \acrshort{5g-advanced} and \gls{6g} architectures, \gls{ns} remains essential for multiplexing heterogeneous services over shared physical infrastructure. However, deploying critical \gls{urllc} alongside standard broadband introduces severe cross-slice vulnerabilities. The implicit assumption that logical slicing inherently guarantees security fails against data plane exploits, as shared \glspl{upf} remain susceptible to resource starvation and protocol tunneling. To address this, we propose the Slice Isolation-as-a-Service framework, which treats topological data plane isolation as an on-demand, monetizable capability. We formulate the multi-\gls{upf} resource allocation problem as a \gls{minlp} model to strictly enforce physical and virtual isolation boundaries based on tenant security tiers, minimizing provisioning costs and end-to-end latency. To validate the framework, we engineered a novel containerized \gls{5g} testbed integrating Free5GC with Nokia Containerlab. Empirical evaluations under active volumetric flooding and the recently discovered \acrshort{gtp}-in-\acrshort{gtp} reflection attacks reveal that shared and \gls{vnf}-only isolation fail catastrophically, suffering up to $30\%$ packet loss and complete bandwidth starvation. Our results prove that \gls{minlp}-driven strict physical isolation successfully neutralizes cross-slice injection and preserves $99.999\%$ reliability for mission-critical slices, confirming our isolation framework as a crucial proactive defense for next-generation core networks.
\end{abstract}





\maketitle

\section{Introduction}
As cellular networks transition from \gls{5g} towards \gls{5g-advanced} and future \gls{6g} systems, \gls{ns} remains a fundamental architectural pillar for supporting heterogeneous services over a common physical infrastructure~\cite{ITU2083,taskou2023end}. Rather than designing an entirely new core from scratch, recent \gls{3gpp} Release 20 standardization studies on the \gls{6g} system architecture (\gls{3gpp} TR 23.801-1~\cite{3gpp.23.801}) establish that \gls{6g} will use the current \gls{5g} architecture as a baseline starting point. Consequently, \gls{6g} will preserve the fundamental \gls{ran} and Core functionality split, reuse or evolve the majority of the existing \gls{5g} \glspl{nf} within the \gls{sba}, and adopt \gls{5g} network slicing as the foundation for next-generation service differentiation~\cite{3gpp.23.801}. 

\subsection{Network Slicing and Resource Multiplexing}
\gls{ns} enables \glspl{mno} to partition a single shared physical infrastructure into multiple independent, logical end-to-end networks, each tailored to the specific \gls{sla} requirements of distinct verticals~\cite{3gpp.23.501, wong20225g}. To maximize hardware utilization and reduce \gls{capex} and \gls{opex}, \glspl{mno} frequently multiplex multiple slices over shared \glspl{vnf} and common physical servers~\cite{Mei2020b, mahyoub2025security}. In the cellular \gls{dp}, this means user traffic from multiple coexisting slices often traverses shared \glspl{upf} encapsulated within \gls{gtp} User Plane (GTP-U) tunnels~\cite{ietf_uplane,3gpp.29.281}. While \glspl{mno} can theoretically provision slices across a spectrum of sharing configurations, ranging from fully shared \glspl{vnf} and physical nodes to partially isolated containers or strictly dedicated hardware, determining the optimal level of isolation requires balancing resource efficiency against cross-slice security risks~\cite{kotulski2017end,wong20225g}.

\subsection{Motivation}
Despite its standardized maturity and a projected \$200 billion market potential by 2030~\cite{ericsson2030}, commercial end-to-end network slicing has seen limited deployment by operators, largely due to slow \gls{5g} Standalone (SA) core rollouts, orchestration complexity, and inter-slice security isolation~\cite{ericsson_emr_2026, 10242032}. This challenge is particularly critical in emerging \gls{5g-advanced} and \gls{6g} private networks, such as smart power grids, automated manufacturing plants, military or healthcare deployments, where mission-critical \gls{urllc} slices coexist alongside less-trusted consumer or third-party \gls{iot} slices~\cite{3gpp.22.104, 11546234}. In such multi-tenant environments, a compromised device in a low-security slice can serve as a launchpad for cross-slice attacks against high-criticality slices~\cite{sattar2025secure, abdulghaffar2024impact}. 

Furthermore, because the majority of cellular network traffic consists of user data plane traffic~\cite{ericsson_emr_2026}, securing and optimizing \gls{dp} resources (specifically \glspl{upf}) is paramount. Since \gls{5g-advanced} and early \gls{6g} architectures will continue to rely on the \gls{5g} \gls{sba} and \gls{gtp}-U tunneling for data plane routing~\cite{3gpp.29.281, 3gpp.23.801, 3gpp.23.501}, the protocol vulnerabilities and hardware resource contention bottlenecks present in current \gls{5g} \glspl{upf} will be directly inherited by \gls{6g} networks if proactive security mechanisms are not implemented.

\subsection{Problem Statement}
Optimizing and deploying coexisting network slices introduces a fundamental architectural conflict between satisfying diverse, often contradicting slice \gls{qos} requirements and guaranteeing robust cross-slice security~\cite{sattar2025secure, souza2024enabling, taskou2023end}. On one hand, mission-critical \gls{urllc} services mandate deterministic, ultra-low latency ($1\text{--}10$~ms) and $99.999\%$ packet reliability~\cite{3gpp.22.104, 3gpp.22.261}, whereas \gls{embb} services prioritize high-throughput data transfer. On the other hand, modern networks face an unprecedented escalation in volumetric and amplification \gls{dos} threats. Recent global threat intelligence from Cloudflare highlights a massive surge in hyper-volumetric network-layer attacks, including reflection vectors and multi-terabit floods exceeding $5.6$ to $31.4$~Tbps, capable of saturating shared network queues within seconds~\cite{cloudflare_ddos}.

Within sliced cellular cores, an implicit assumption is that logical slice separation inherently provides security isolation~\cite{kotulski2017end}. However, when underlying physical servers, hypervisors, or \gls{upf} instances are shared across slices, this logical boundary breaks down:

\begin{itemize}
    \item \textbf{Protocol-Level Cross-Slice Bleeding:} Standard \gls{upf} lacks recursive inner-header verification during \gls{gtp}-U decapsulation, enabling adversaries to execute cross-slice \gls{gtp}-in-\gls{gtp} reflection attacks~\cite{shaik2025uncovering} that inject spoofed traffic into adjacent slices, causing severe quota exhaustion and billing fraud. 
    \item \textbf{Hardware-Level Resource Starvation:} High-rate volumetric flooding attacks~\cite{abdulghaffar2024impact, husseinat2025study} launched from an adjacent slice can exhaust shared host CPU cycles, memory bandwidth, and virtual Layer-2 bridges, violating deterministic \gls{urllc} \glspl{sla} even when slices are assigned separate \gls{vnf} containers.
\end{itemize}
In contrast, blindly provisioning dedicated physical servers for every network slice eliminates the multiplexing benefits of virtualization and is financially unsustainable for \glspl{mno}~\cite{wong20225g}. Thus, operators lack a unified mechanism to dynamically match slice isolation depth to tenant criticality while minimizing infrastructure costs.

\subsection{Proposed Approach: Slice Isolation-as-a-Service}
To resolve the trade-off between strict security isolation and resource efficiency, we propose the \textbf{Slice Isolation-as-a-Service} framework. Our approach transforms \gls{dp} isolation from a static configuration into a tiered, on-demand, and monetizable capability. Under this framework, tenants request one of three security tiers (\textit{Low}, \textit{Medium}, or \textit{High}) during slice registration based on their operational criticality and willingness to pay.

We realize this framework through a rigorous two-stage methodology combining mathematical optimization with empirical testbed validation: \textcircled{1} We formulate the multi-\gls{upf} \gls{dp} placement and routing problem as a \gls{minlp}. The solver dynamically computes topologies that jointly minimize \gls{vnf} processing capacity, physical node baseline overhead, and end-to-end traffic latency while enforcing strict mathematical constraints that govern physical and virtual resource sharing across the three tiers. \textcircled{2} Rather than relying solely on theoretical simulations, we translate the optimal topologies generated by our \gls{minlp} solver into a custom multi-node \gls{5g} testbed orchestrating Free5GC~\cite{free5gc} and Nokia Containerlab~\cite{containerlab}. We then subject these topologies to live cross-slice \gls{gtp}-in-\gls{gtp} reflection~\cite{shaik2025uncovering} and volumetric flooding attacks~\cite{abdulghaffar2024impact} to empirically verify their resilience across transport- and network-layer metrics.

\subsection{Contributions}
To address the critical security gaps in current \gls{5g-advanced} and future \gls{6g} data plane architectures, this paper makes the following primary contributions:
\begin{enumerate}
    \item \textbf{Formulation of the Slice Isolation-as-a-Service Framework:} To proactively secure \gls{dp} traffic, we introduce the novel Slice Isolation-as-a-Service framework, formalized as a \gls{minlp} resource allocation model. The model mathematically enforces strict physical and virtual isolation boundaries based on tenants' required security tiers, dynamically optimizing \gls{vnf} placement to satisfy strict \gls{sla} performance requirements while minimizing resource provisioning costs and \gls{opex} for \glspl{mno} \textbf{(Section~\ref{sec:Model})}.
    \item \textbf{Novel Containerized \gls{5g} Testbed:} Bridging the gap between theoretical optimization and practical deployment, we engineered a highly flexible, multi-node \gls{5g} testbed using open-source tools. By uniquely integrating Free5GC \glspl{upf} with Nokia Containerlab, we overcame standard kernel namespace limitations to orchestrate a distributed, multi-interface \gls{upf} data plane capable of seamlessly hosting optimization-driven architectural topologies and processing actual generated network traffic for evaluation~\textbf{(Section~\ref{sec:testbed})}.
    \item \textbf{Demonstration of the Effectiveness of the Proposed Framework:} This is done by performing the following two sub-contributions:
    \begin{enumerate}
        \item \textbf{Demonstration of Cross-Slice \acrfull{dp} Susceptibility to Known Attacks:} We challenge the implicit assumption that logical network slicing inherently guarantees security isolation~\cite{kotulski2017end}. By successfully implementing a newly discovered \acrshort{gtp}-in-\acrshort{gtp} reflection attack (ACM CCS 2025) \cite{shaik2025uncovering} and common volumetric flooding \cite{abdulghaffar2024impact} attacks within a sliced environment, we empirically prove that shared data plane infrastructure, specifically \glspl{upf}, is fundamentally susceptible to cross-slice bleeding, latency degradation, and billing fraud \textbf{(Section~\ref{sec:results_gtp} and \ref{sec:results_flood})}.
        \item \textbf{Mitigation of Attacks using the Proposed Approach:} We extensively evaluate the efficacy of the proposed Slice Isolation-as-a-Service model by deploying its mathematically generated network topologies directly into the \gls{5g} testbed \textbf{(Section~\ref{sec:exp_topology})}. Under active \gls{dp} attacks, specified in Contribution 3(a), our empirical results validate that the model's secure allocation strategy effectively eliminates cross-slice susceptibility, successfully preserving $99.999\%$ packet reliability and deterministic latency for mission-critical slices \textbf{(Section~\ref{sec:results_gtp} and \ref{sec:results_flood})}.
    \end{enumerate}
\end{enumerate}

The remainder of the paper is structured as follows. Section~\ref{sec:related} reviews related work on network slice resource allocation and security. Section~\ref{sec:threat} defines the threat model, assumptions, and adversary capabilities. Our research methodology, the proposed Slice Isolation-as-a-Service solution, and the evaluated cross-slice attacks are explained in Section~\ref{sec:methodology}. Section~\ref{sec:Model} details the mathematical formulation of the \gls{minlp} optimization model. The design and implementation of the containerized \gls{5g} security testbed is described in Section~\ref{sec:testbed}, while Section~\ref{sec:exp_topology} outlines the optimization-driven experimental topologies and evaluation metrics. Section~\ref{sec:results_gtp} and Section~\ref{sec:results_flood} analyze the empirical testbed results under \gls{gtp}-in-\gls{gtp} reflection and flooding attacks, respectively. Section~\ref{sec:perf_eval} evaluates the computational scalability and resource overhead of the optimization model. Section~\ref{sec:limitation} discusses the limitations of our work. Section~\ref{sec:ethical} outlines the ethical considerations of our experimental methodology, and finally, Section~\ref{sec:conclusion} concludes the paper.

\section{Related Work} \label{sec:related}
This section reviews recent literature on network slice resource allocation, security-aware slice isolation models, and empirical \gls{5g} slicing testbeds. Table~\ref{tab:related_work} summarizes the most closely related studies and highlights how the proposed Slice Isolation-as-a-Service framework advances the state of the art.

\subsection{Slice Resource Allocation and Coexistence}

Several studies model resource allocation and \gls{vnf} sharing to support coexisting slices while minimizing deployment costs. Souza~\textit{et al.}~\cite{souza2024enabling} design a Continuous-Time Markov Chain (CTMC) queuing model to analyze container auto-scaling, setup delays, and availability for coexisting \gls{embb} and \gls{urllc} slices within a single \gls{mec}-\gls{nfv} node. Taskou~\textit{et al.}~\cite{taskou2023end} formulate a multi-objective \gls{minlp} model, solved using convex approximation, that jointly slices \gls{ran} and \gls{cn} resources for \gls{embb} and \gls{urllc} slices to minimize end-to-end (E2E) energy consumption and resource costs. Focusing on the \gls{5g} user plane, Leyva-Pupo~\textit{et al.}~\cite{Leyva-Pupo2022} propose an \gls{ilp} model and heuristic-based solutions to optimize multi-branch \gls{upf} placement and chaining in \gls{mec}. Similarly, Mei~\textit{et al.}~\cite{Mei2020b} formulate a slice embedding model with shareable \glspl{vnf} to maximize slice acceptance ratios, while Zharabad~\textit{et al.}~\cite{JALALIAN2023103631} combine an \gls{ilp} model with deep learning request prediction to minimize \gls{vnf} sharing costs in virtualized \gls{5g} cores. 

\textit{Comparison with the current work:} While the studies in this subsection~\cite{souza2024enabling, taskou2023end, Leyva-Pupo2022, Mei2020b, JALALIAN2023103631} effectively optimize \gls{qos}, latency, and resource utilization, they assume benign traffic environments and permit unrestricted \gls{vnf} sharing up to hardware capacity limits. None of these works account for cross-slice security threats or validate their generated topologies on an empirical \gls{5g} testbed.

\begin{table}[t]
    \centering
    \caption{Comparison of Related Work in 5G and Beyond Network Slicing}
    \label{tab:related_work}
    \resizebox{\textwidth}{!}{%
    \begin{tabular}{l|c|c|c|c|c|c|c|c} \toprule[1.5pt]
        \rowcolor{gray!20}
        \textbf{Ref.}       & \textbf{Mathematical} & \textbf{Multi-UPF}    & \textbf{Security}     & \textbf{VNF}      & \textbf{Physical Host}    & \textbf{Empirical 5G} & \textbf{Volumetric Flood} & \textbf{Cross-Slice GTP-in-GTP} \\
        \rowcolor{gray!20}
                            & \textbf{Optimization} & \textbf{Data Plane}   & \textbf{Constraints}  & \textbf{Isolation} & \textbf{Isolation}       & \textbf{Core Testbed} & \textbf{Evaluation} & \textbf{Reflection Evaluation} \\ \midrule[1.5pt] 
\cite{souza2024enabling}    & $\times$ (CTMC)       & $\times$              & $\times$              &  $\times$          & $\times$             & $\times$ (Simulation)   & $\times$              & $\times$ \\
        \rowcolor{blue!7}
        \cite{taskou2023end} & \checkmark (MINLP)   & $\times$              & $\times$              & $\times$          & $\times$                  & $\times$ (Simulation)     & $\times$              & $\times$ \\
        \cite{Leyva-Pupo2022} & \checkmark (ILP)    & \checkmark (I/MI/A-UPF) & $\times$            & $\times$           & $\times$                 & $\times$ (Simulation) & $\times$                  & $\times$ \\
        \rowcolor{blue!7}
        \cite{Mei2020b}     & \checkmark (ILP)      & $\times$              & $\times$              & $\times$            & $\times$              & $\times$ (Simulation) & $\times$              & $\times$ \\
    \cite{JALALIAN2023103631} & \checkmark (ILP/DL) & $\times$              & $\times$              & $\times$           & $\times$              & $\times$ (Simulation) & $\times$ &            $\times$ \\ \midrule[1.25pt]
        \rowcolor{blue!7}
        \cite{wong20225g} & \checkmark (LP)         & $\times$              & \checkmark (Min.\ Sec.)     & \checkmark (Logical) & \checkmark (Bare-Metal) & $\times$ (Analytical) & $\times$ & $\times$ \\
        \cite{Li8869775} & \checkmark (ILP)         & $\times$              & \checkmark (Node Trust) & $\times$             & \checkmark (Host Mapping)  & $\times$ (Simulation) & $\times$ & $\times$ \\
        \rowcolor{blue!7}
        \cite{Coelho9627736} & \checkmark (MILP)    & $\times$              & $\times$ (Static Policy) & $\times$        & $\times$             & $\times$ (Simulation)       & $\times$      & $\times$ \\
        \cite{Javadpour10064361} & \checkmark (RL) & $\times$                 & \checkmark (DDoS Defense) & $\times$ & \checkmark (Host Isolation)        & $\times$ (Simulation)     & \checkmark (Simul.\ DDoS) & $\times$ \\
        \rowcolor{blue!7}
        \cite{11546234} & \checkmark (NSGA-II/DQN) & $\times$               & \checkmark (GCN Policy)       & \checkmark (Micro-Seg.) & $\times$ & $\times$ (Simulation) & $\times$ & $\times$ \\
        \cite{sattar2025secure} & \checkmark (MILP) & $\times$               & \checkmark (ISI) & $\times$   & \checkmark (Host ISI) & $\times$ (OpenVZ) & \checkmark (Tcpreplay) & $\times$ \\
        \rowcolor{blue!7}
        \cite{mahyoub2025security} & \checkmark (MINLP)     & $\times$      & \checkmark (Traffic/Exposure) & \checkmark (Critical VNFs) & $\times$ & $\times$ (JuMP Simulation) & $\times$ & $\times$ \\\midrule[1.25pt]
        \cite{grings2025nasp} & $\times$  & $\times$ (Single UPF)     & $\times$  & \checkmark (K8s/VLAN) & $\times$                               & \checkmark (Free5GC/K8s)  & $\times$      & $\times$ \\
        \rowcolor{blue!7}
        \cite{chai2021} & $\times$              & $\times$ (Single UPF)     & $\times$              & \checkmark (Dedicated NFs)    & $\times$    & \checkmark (Free5GC)       & $\times$      & $\times$ \\
        \cite{amponis2022threatening} & $\times$    & $\times$ (Single UPF)     & $\times$              & $\times$              & $\times$              & \checkmark (Open5GS)          & $\times$       & $\times$ \\
        \rowcolor{blue!7}
        \cite{Khan10056693} & $\times$         & $\times$ (Single UPF)     & $\times$          & $\times$          & $\times$              & \checkmark (Free5GC)      & \checkmark (DP DDoS)          & $\times$ \\
    \cite{abdulghaffar2024impact} & $\times$            & $\times$ (Single UPF)         & $\times$      & \checkmark (SMF)      & \checkmark (UPF)        & \checkmark (Free5GC)  & \checkmark (Ping/Reg.) & $\times$ \\
        \rowcolor{blue!7}
        \cite{husseinat2025study} & $\times$            & $\times$ (Single UPF)             & $\times$     & \checkmark (SMF)   & \checkmark (UPF)   & \checkmark (Free5GC)  & \checkmark (Ping/UDP/Reg.) & $\times$ \\ \midrule[1.25pt]
        \textbf{This Work} & \checkmark \textbf{(MINLP)} & \checkmark \textbf{(I/C/P-UPF)} & \checkmark \textbf{(3 Isolation Tiers)} & \checkmark \textbf{(Med./High)} & \checkmark \textbf{(High Tier)} & \checkmark \textbf{(Free5GC+Clab)} & \checkmark \textbf{(TRex 10k PPS)} & \checkmark \textbf{(Spoofed IP/TEID)} \\ 
        \bottomrule[1.25pt]
    \end{tabular}
    }
\end{table}

\subsection{Security-Aware Slice Isolation Models}
To mitigate multi-tenant risks, several works incorporate security isolation into slice provisioning models. Wong~\textit{et al.}~\cite{wong20225g} introduced an eight-layer Network Slice as a Service (NSaaS) security taxonomy and analytically modeled the economic trade-offs between protocol-stack isolation points, tenant control levels, and deployment costs. Li~\textit{et al.}~\cite{Li8869775} formulate secure \gls{5g} core slice provisioning as an \gls{ilp} model with a multi-criteria decision-making method, whereas Coelho~\textit{et al.}~\cite{Coelho9627736} optimize slice provisioning across \gls{ran} functional splits and control/data-plane separations using predefined \gls{vnf} sharing policies. To counter \gls{ddos} threats, Javadpour~\textit{et al.}~\cite{Javadpour10064361} propose a reinforcement learning-based slice isolation model to improve request acceptance in Beyond \gls{5g} networks, while Zhang~\textit{et al.}~\cite{11546234} combine \gls{nsga}, Deep Q-Networks (DQN), and \gls{gcn} to adjust application-layer micro-segmentation in \gls{5g} power private networks. 

In our prior work, Sattar~\textit{et al.}~\cite{sattar2025secure} formulated a proactive \gls{milp} model enforcing hardware inter-slice isolation (ISL), intra-slice \gls{vnf} separation, and secure hypervisor zones to mitigate \gls{ddos} attacks on an OpenVZ virtualization testbed. Building on this, Mahyoub ~\textit{et al.}~\cite{mahyoub2025security} proposed a \gls{5g} \gls{cp} procedure-level \gls{minlp} model that minimizes total \gls{vnf} capacity and delay across standard \gls{3gpp} \gls{cp} procedures by limiting \gls{vnf} traffic and isolating ingress \glspl{vnf} exposed to untrusted access networks.

\textit{Comparison with the current work:} Existing security models in the literature either rely on high-level protocol abstractions~\cite{wong20225g}, require sharing policies to be manually predefined as static inputs~\cite{Coelho9627736}, or employ heuristic and multi-criteria approximations~\cite{11546234, Li8869775, Javadpour10064361}; crucially, none of these studies validate their models on an empirical \gls{5g} core testbed~\cite{wong20225g, Coelho9627736, 11546234, Li8869775, Javadpour10064361}. Furthermore, while our prior work evaluated host-level isolation on a custom three-node authentication service~\cite{sattar2025secure} and \gls{cp} procedure sharing in simulation~\cite{mahyoub2025security}, the present study substantially extends our \gls{minlp} formulation~\cite{mahyoub2025security} to the \gls{5g-advanced} and \gls{6g} \textit{data plane}. Specifically, we model \gls{3gpp}-compliant multi-\gls{upf} chains ($\text{I-\gls{upf}}\rightarrow\text{C-\gls{upf}}\rightarrow\text{P-\gls{upf}}$), enforce three-tier physical and virtual Slice Isolation-as-a-Service isolation constraints, and deploy the resulting topologies on a containerized \gls{5g} testbed.

\subsection{Empirical Slicing Platforms and Attack Evaluations}
Recent experimental studies have utilized open-source cellular platforms to evaluate slice orchestration and attack impacts. Grings~\textit{et al.}~\cite{grings2025nasp} present the Network slice as a Service Platform (NASP), which translated GSMA templates into E2E slices across Kubernetes, Free5GC, and Mininet using \gls{vlan} segmentation. Similarly, Chai and Lin~\cite{chai2021} deployed an OpenStack and Free5GC testbed to benchmark registration delay and throughput across dedicated \gls{smf} and \gls{upf} configurations. Under adversarial conditions, Amponis~\textit{et al.}~\cite{amponis2022threatening} demonstrate N4 \gls{pfcp} \gls{dos} attacks against \gls{uav} sessions in an Open5GS testbed, while Khan~\textit{et al.}~\cite{Khan10056693} evaluate \gls{dp} \gls{ddos} attacks in a Free5GC and UERANSIM testbed and train a deep learning detector \textit{SliceSecure}. In our preliminary testbed studies, AbdulGhaffar~\textit{et al.}~\cite{abdulghaffar2024impact} and Husseinat~\textit{et al.}~\cite{husseinat2025study} evaluated shared versus dedicated \gls{smf} and \gls{upf} configurations under \gls{icmp} ping, \gls{udp}, and registration floods, demonstrating severe degradations in \gls{tcp} throughput and \gls{rtt}~\cite{abdulghaffar2024impact} as well as \gls{xr} frame burstiness~\cite{husseinat2025study} when \glspl{upf} are shared. Recently, Shaik~\textit{et al.}~\cite{shaik2025uncovering} uncovered the \gls{gtp}-in-\gls{gtp} protocol tunneling vulnerability, demonstrating how nested \gls{gtp} payloads bypass \gls{upf} routing boundaries. The authors evaluated six \gls{5g} core networks, including, Free5GC~\cite{free5gc}, Open5GS~\cite{Open5gs}, OpenAirInterface-\gls{5g}~\cite{OAI}, SD-Core~\cite{sd-core}, and two commercial \gls{5g} networks.

\textit{Comparison with the current work:} Prior testbed studies either evaluate slice orchestration under benign traffic~\cite{grings2025nasp, chai2021}, target \gls{cp} \gls{pfcp} signaling~\cite{amponis2022threatening}, or rely on reactive anomaly detection~\cite{Khan10056693}. Moreover, our earlier empirical works~\cite{abdulghaffar2024impact, husseinat2025study} tested single-\gls{upf} topologies without an underlying optimization model and did not separate \gls{vnf} containers from physical host isolation. On the other hand, Shaik~\textit{et al.}~\cite{shaik2025uncovering} evaluated \gls{gtp}-in-\gls{gtp} exploits only in unsliced cores. To the best of our knowledge, this work is the first to integrate a proactive \gls{minlp} multi-\gls{upf} optimization model with a containerized multi-node \gls{5g} testbed to empirically defend against both cross-slice \gls{gtp}-in-\gls{gtp} reflection and volumetric flooding attacks across physical and virtual isolation tiers.

\section{Threat Model} \label{sec:threat}
We explain the threat model by outlining the assumptions and adversary capabilities in the scope of this study.

\subsection{Assumptions}
To evaluate the resilience of the proposed solution, the following assumptions are made:
\begin{itemize}
    \item The target \gls{5g} network is a virtualized architecture that supports running multiple network slices by either sharing or isolating the \glspl{vnf}.
    \item The attacker \gls{ue} is a legitimate, authenticated user of the network who has successfully completed standard registration and \gls{pdu} session establishment procedures to obtain a valid \gls{ip} address and attach to its specific network slice.
    \item The attacker cannot physically modify the bare-metal hardware and hypervisor configurations of the system hosting \gls{5g} \glspl{vnf}.
    \item The attack scope is strictly limited to the \gls{dp}. The \gls{5g} \gls{cp} functions, such as the \gls{amf} and \gls{smf}, are assumed to be fully protected and cannot be compromised by signaling attacks.
\end{itemize}

\subsection{Adversary Capabilities}
The attacker is assumed to possess the following knowledge and technical capabilities:

\begin{itemize}
    \item The attacker has a working knowledge of \gls{3gpp} \gls{5g} specifications and is proficient in attack methodologies. 
    \item The adversary possesses the necessary programming capabilities to craft protocol-compliant \gls{ip} packets.
    \item The attacker is capable of injecting specially crafted payloads directly into the uplink data stream, forcing the \gls{gnb} to encapsulate them within \gls{gtp} User Plane (GTP-U) headers and forward them blindly through the \gls{upf}. Crucially, the attacker has the capability to spoof internal \gls{ip} addresses and Victim \glspl{teid} within these crafted payloads to exploit decapsulation vulnerabilities at the \gls{upf}.
    \item Through reconnaissance, public documentation, or probing, such as $traceroute$, the attacker can successfully discover the internal \gls{ip} addresses of critical core components, including the \gls{upf}.
    \item The attacker has the infrastructure or tools required to generate a high volume of traffic (e.g., volumetric \gls{udp}/\gls{icmp} floods) from a single compromised \gls{ue}, specifically designed to exhaust the computational resources of the operator's \glspl{vnf}.
   \item By combining volumetric generation with protocol manipulation, the attacker aims to exploit shared \glspl{vnf} to maliciously degrade the performance and service availability of legitimate \glspl{ue} operating in adjacent network slices.  
\end{itemize}

\section{Research Methodology and Proposed Solution} \label{sec:methodology}
\subsection{Research Methodology}
Our methodology involves a two-step approach: first, a mathematical optimization model calculates secure resource allocations based on slice-specific isolation requirements; second, the resulting topologies are deployed and empirically evaluated on the realistic \gls{5g} testbed under active data plane attacks.

\begin{figure}[t]
    \centering
    \includegraphics[width=\linewidth]{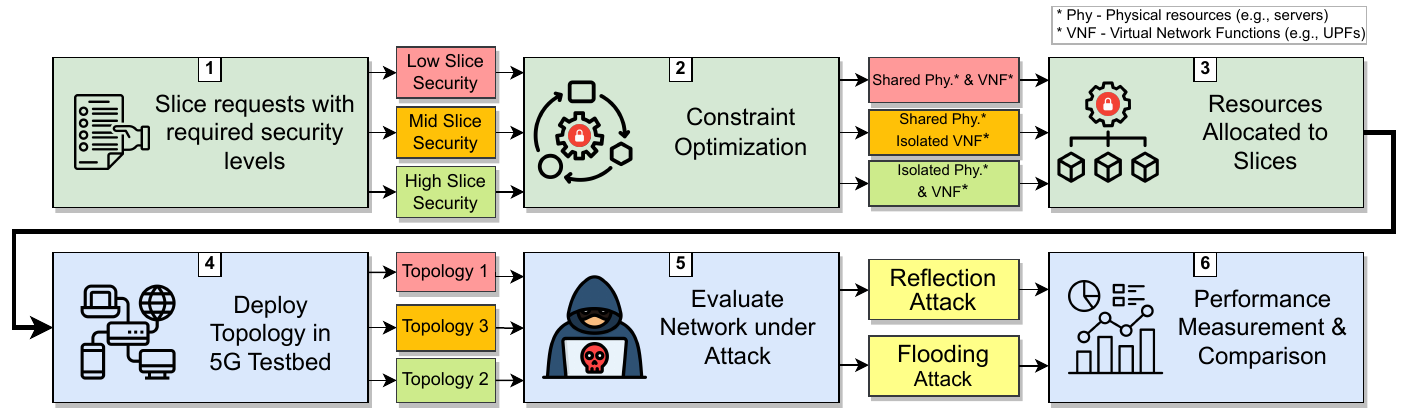}
    \caption{Research Methodology. Green blocks on the top represent the optimization model methodology, and Blue blocks show the testbed-based empirical experimentation.}
    \label{fig:Methodology}
\end{figure}

Figure~\ref{fig:Methodology} presents an overview of our methodology. The green blocks at the top of the figure summarize the methodology of the optimization model. The proposed `Slice Isolation-as-a-Service' optimization model is explained in Section~\ref{sec:Model}. Initially, multiple slice initialization requests with different security requirements are provided to the model as input. Each slice may request one of three security tiers based on traffic criticality. As summarized in Table~\ref{tab:Security_levels}, these security levels dictate the physical and virtual resource sharing constraints:

\begin{itemize}
    \item \textbf{Low Security:} \glspl{vnf} and underlying physical servers are fully shared among slices to maximize resource utilization and minimize deployment cost. This configuration is suitable for standard consumer \gls{embb} applications.
    \item \textbf{Medium Security:} Slices are allocated dedicated \gls{vnf} instances (container-level isolation) but share underlying physical compute nodes. This is suited for standard enterprise communications.
    \item \textbf{High Security:} Slices are granted dedicated \gls{vnf} instances running on strictly isolated physical nodes. This configuration is mandated for mission-critical \gls{urllc} applications (e.g., industrial automation, autonomous transport, smart grid control) requiring guaranteed determinism and immunity from adjacent slice activity.
\end{itemize}

These slice initialization requests are then passed to the proposed constraint optimization model, which allocates the required network resources to each slice according to its security requirements. We call these allocation configurations `Topologies'.

After obtaining the topologies from the optimization model, we deploy them in the \gls{5g} testbed. This step is shown in blue in Figure~\ref{fig:Methodology}. Further details on our \gls{5g} testbed are provided in Section~\ref{sec:testbed}. Each of the three implemented topologies, corresponding to Low, Medium, and High slice security configurations, is then evaluated under Reflection attack and Flooding attack scenarios. The impact of attacks on user traffic is evaluated using various performance metrics.

\begin{table}[!t]
    \centering
    \caption{Slice Security Levels, their Resource Sharing Levels, and Target Use Cases}
    \begin{tabular}{c||c||c||c} \toprule[1.5pt]
        \rowcolor{blue!7}
        \textbf{Security level} & \textbf{Physical Resources} & \textbf{Virtual Resources} & \textbf{Target Use Cases}\\ \midrule[1.5pt] 
        Low Security& Shared & Shared & Consumer \gls{embb}\\ 
        \rowcolor{blue!7}
        Medium Security& Shared & Isolated & Non-critical \gls{iot}\\ 
        High Security& Isolated & Isolated & Mission-Critical \gls{urllc}\\ 
        \bottomrule[1.25pt]
    \end{tabular}
    \label{tab:Security_levels}
\end{table}

\subsection{Proposed Solution: Slice Isolation-as-a-Service}

To address cross-slice susceptibility without permanently over-provisioning dedicated hardware for every tenant, we propose the `\textbf{Slice Isolation-as-a-Service}' framework (Section~\ref{sec:Model}). Under this framework, physical and virtual isolation is treated as an on-demand, monetizable service feature rather than a static network-wide property. Industry forecasts highlight that network slicing represents a $\$200$ billion market opportunity by 2030, driven primarily by enterprise (\gls{b2b2x}) demand for highly reliable, application-specific networks \cite{ericsson2030}. To capture this value, telecommunications vendors emphasize transforming \gls{5g} into a services and differentiation platform, where operators can monetize tiered service models by guaranteeing strict \glspl{sla} for premium pricing \cite{nokia_whitepaper}. By allowing tenants to declare their required security isolation level during slice registration, our orchestration framework dynamically guarantees the requested level of topological separation, allowing operators to monetize strict physical isolation for mission-critical use cases.

The concept of on-demand isolation aligns closely with recent standardization efforts by the \gls{ietf} in RFC 9543 \cite{rfc9543}. The \gls{ietf} explicitly establishes the ``Isolation as a Service Requirement'' (in RFC 9543 Section 8.1), defining isolation as a formal capability that customers can request to ensure that external network behaviors and adjacent services have no negative impact on their service delivery. While the \gls{ietf} framework provides a broad architectural blueprint for achieving this separation through various underlying resource partitioning techniques, such as L2/L3 \glspl{vpn}, our proposed `\textbf{Slice Isolation-as-a-Service}' framework offers a concrete, optimization-driven realization of this requirement specifically tailored for the \gls{5g} and Beyond Core \acrfull{dp} (\gls{upf}).

By combining mathematical \gls{minlp} with automated containerized deployment, our framework provides the rigorous mathematical engine necessary to translate the \gls{ietf}'s high-level isolation requirements into strictly enforced physical and virtual boundaries within the cellular core. The complete mathematical formulation of this \gls{minlp} optimization model is detailed in Section~\ref{sec:Model}.

\subsection{Cross-Slice Attacks Evaluated in the 5G Testbed} \label{sub:attacks}
We implement the following two attacks to evaluate their impact on various slice isolation configurations in the \gls{5g} testbed:

\begin{figure}[t]
    \centering
    \includegraphics[width=1\linewidth]{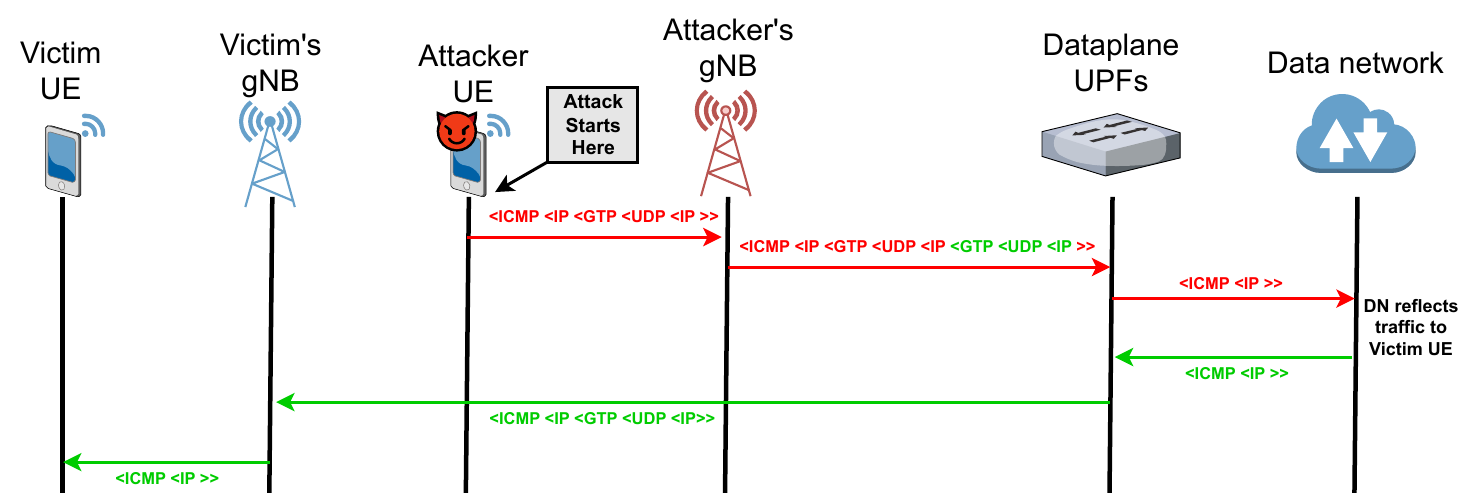}
    \caption{Reflection Attack Sequence Diagrams. \textcolor{red}{Red arrows for attack traffic.} \textcolor{Green}{Green is legitimate traffic, but becomes part of the attack}. \textbf{Note:} The attacker's gNB is not under the control of the attacker, but instead it is a legitimate gNB serving the attacker (malicious user)~\cite{shaik2025uncovering}.}
    \label{fig:reflection_call_flow}
\end{figure}

\begin{itemize}
    \item \textbf{Reflection Attack:} This attack exploits the lack of nested \gls{gtp} header verification and routing validation within standard \gls{upf} implementations to inject traffic into a victim \gls{ue}'s uplink \gls{gtp} tunnel \cite{3gpp.29.281, mahyoub2024security}. Recently uncovered in the context of \gls{5g} by Shaik \textit{et al.} at ACM \gls{ccs} 2025~\cite{shaik2025uncovering}, the attacker initially crafts a malformed \gls{gtp}-in-\gls{gtp} packet. As detailed in Figure~\ref{fig:GTP_packet} in Appendix~\ref{sec:appendix}, the inner \gls{ip} and \gls{gtp} headers are maliciously constructed using the victim \gls{ue}'s \gls{ip} address and \gls{teid}. The attacker sets the destination target to the victim's \gls{upf} in the inner \gls{ip} header and \gls{gtp} header. The inner payload (e.g., an \gls{icmp} or \gls{dns} request) is directed to a public server.
        
    The malformed packet is then sent normally by the attacker \gls{ue} to their connected \gls{gnb}. The \gls{gnb} receives the packet and encapsulates it with a legitimate outer \gls{gtp} header containing the information (\gls{teid}, \gls{ip}) of the attacker \gls{ue} and sends it to the data plane \gls{upf}. Upon reception, the \gls{upf} removes the outer \gls{gtp} header and blindly routes the inner payload based on the spoofed \gls{teid}-\gls{ip} pair inside the malformed \gls{gtp} header. Since the inner \gls{teid}-\gls{ip} pair is meticulously crafted by the attacker to target a victim \gls{ue}, the \gls{upf} sends the inner payload from the victim's uplink tunnel, appearing to be originated from the victim \gls{ue}. Consequently, the public server's response is routed back through the victim’s downlink tunnel. This effectively consumes the victim’s resources in both directions, leading to severe billing fraud and quota depletion; we highlight these impacts in detail in Section~\ref{sec:results_gtp}. The sequence of the reflection attack is shown in Figure~\ref{fig:reflection_call_flow}.

    \item \textbf{Data plane Flooding Attack:} While the reflection attack targets logical routing vulnerabilities, the volumetric flooding attack targets the physical processing limitations of the shared infrastructure. To execute a flooding attack, legitimate \gls{icmp} requests from the attacker \gls{ue} directed at the \gls{dn} are captured in a \gls{pcap} file on the \gls{gnb}'s N3 interface to obtain valid, session-specific \gls{gtp}-U headers. To circumvent emulation bottlenecks (detailed in Section~\ref{sec:testbed_challenges}), this captured traffic is replayed directly into the \gls{upf} using the TRex traffic generator \cite{trex} at a sustained rate of 10,000 \gls{pps}. As the \gls{dn} replies with corresponding \gls{icmp} response packets, the bidirectional volumetric flood rapidly exhausts the computational CPU cycles, memory bandwidth, and interface queues of the targeted \glspl{upf}. This attack is utilized to test whether volumetric congestion in one slice can cascade and degrade the performance of user traffic in adjacent slices under varying levels of physical and virtual resource isolation.

\end{itemize}

\section{Slice Isolation-as-a-Service Framework and Problem Formulation} \label{sec:Model}
The proposed resource allocation model builds upon our preliminary formulation introduced in Mahyoub \textit{et al.}~\cite{mahyoub2025security}. However, while our prior work focused primarily on optimizing \gls{vnf} sharing at the \acrfull{cp} procedure level by limiting \gls{vnf} exposure to untrusted access networks, this study substantially extends the mathematical model to formally define and enforce the Slice Isolation-as-a-Service paradigm. We formulate this proactive security resource allocation problem as \acrfull{minlp}. The primary objective of the proposed model is to minimize the total processing capacity required by the \glspl{vnf}, the base provisioning overhead of active physical nodes, and the end-to-end traffic latency, while strictly adhering to isolation constraints demanded by the Slice Isolation-as-a-Service framework.

	\subsection{Data Plane Modeling and Mathematical Notations}

    Let $\mathcal{S}$ denote the set of all network slice requests. Each slice $s \in \mathcal{S}$  accommodates a distinct set of user traffic types, denoted by $\mathcal{T}_s$. To represent the \gls{pdu} session data path required for each traffic type, we model the virtual network as a collection of directed graphs. Specifically, a traffic type $t \in \mathcal{T}_s $ is modeled as $(\mathcal{V}^s_t, \mathcal{R}_t^s)$, where $\mathcal{V}^s_t$ represents the set of required \glspl{vnf}, which in this architecture exclusively represent distributed \gls{5g} \glspl{upf}, and $\mathcal{R}_t^s$ defines the virtual links interconnecting them. Furthermore, each traffic type $t$ demands a specific packet arrival rate $\lambda_t^s~$ and enforces a strict maximum tolerated latency boundary $\delta_t^{s,max}$.

    The broader set of all available \gls{vnf} types is defined as $\mathcal{V}$. To capture the computational behavior of these network functions, each \gls{vnf} type $v \in \mathcal{V}$ is mathematically characterized by the tuple  $ \langle v_{i},~I_{v},~\delta^n_{v_i},~\zeta^{max}_{v},~\mu_{v},~\omega_{v} \rangle $. In this tuple, $v_{i}$ identifies the $i^{th}$ active instance of \gls{vnf} type $v$, while $I_{v}$ contains the complete set of instances for that type distributed across the physical infrastructure. The parameter $\delta^n_{v_i}$ defines the processing delay incurred by instance $v_i$ when hosted on physical node $n$. The hardware limits and performance metrics are defined by $\zeta^{max}_{v}$, the absolute maximum processing capacity limit for an instance of type $v$, $\mu_{v}$, which is the computational capacity required to process a single unit of traffic, and $\omega_{v}$, the maximum number of data units the \gls{vnf} can process per unit of time.

    Finally, the underlying physical \gls{5g} infrastructure is modeled as a directed graph $\mathcal{G}=(\mathcal{N}, \mathcal{L})$. Here, $\mathcal{N}$ represents the set of available physical compute nodes (e.g., bare-metal servers hosting the \gls{5g} core), and $\mathcal{L}$ represents the set of physical communication links connecting them. Each node $n \in \mathcal{N}$ is bounded by a finite maximum processing capacity, $C^{max}_{n}$. Similarly, any physical link $(n,m) \in \mathcal{L}$ connecting node $n$ and node $m$ is constrained by a maximum bandwidth capacity $b{(n,m)}$ and incurs a deterministic propagation delay $d{(n,m)}$, which is proportional to the physical distance of the link. A comprehensive summary of all mathematical notations and variables utilized in this formulation is provided in Table \ref{tab:analyticalSymbols}.

 \begin{table}[tb!]
	\centering
	\caption{Used notations summary}
	\label{tab:analyticalSymbols}
	\begin{tabular}{m{2cm} m{12cm} }\toprule[1.5pt]
    \rowcolor{blue!7}
		\textbf{Parameter} & \textbf{Definition}\\
		\midrule[1.5pt] 
		$\mathcal{S}$& Set of all network slice requests \\
		$\mathcal{T}$& Set of all supported traffic types \\
		$\mathcal{V}$& Set of available \gls{vnf} (\gls{upf}) types \\
		$\mathcal{N}$& Set of physical compute nodes within the infrastructure\\ 
		$\mathcal{L}$& Set of physical links interconnecting the compute nodes in $\mathcal{N}$ \\ 
        $\mathcal{T}_s$& Set of all traffic types demanded by slice $s$ \\
        $\mathcal{R}$ & Set of all virtual links\\
        $\mathcal{R}_t^s$& Set of virtual links configuring the data path for traffic $t$ in slice $s$\\ 
        $\mathcal{V}^s_t $& Set of required \glspl{vnf} for traffic type $t$ within slice $s$ \\
        $I_{v}$ & Set of all instances for \gls{vnf} type $v$	\\        
		$(n,m)$& Directed physical link starting at node $n$ and terminating at node  $m$\\
        $b{(n,m)}$ & Maximum bandwidth capacity of the physical link $(n,m)$ \\
        $d{(n,m)}$ & Propagation delay over physical link $(n,m)$ \\
        $P_n^B$ & Baseline computational overhead to provision a physical node $n$ \\
		$C^{max}_{n}$ & Finite processing capacity bound of node $n$		\\
		$v_{i}$ & The $i^{th}$ active instance of \gls{vnf} type $v$ \\
		$(v_i,z_j)$& The virtual link connecting instance $v_i$ to $z_j$  \\
        $\zeta^{max}_{v}$ & Absolute maximum processing capacity limit for a \gls{vnf} instance of type $v$ \\
        $\mu_{v}$ & Computational capacity required by \gls{vnf} $v$ to process one traffic unit	\\
		$\omega_{v}$&Number of processed data units per time interval by the \gls{vnf} type $v$ 	\\
		$\lambda_t^s$& Required packet arrival rate for traffic type $t$  in slice $s$\\
        $\delta_t^s$ & The delay experienced by the traffic type $t$ of slice $s$ \\
		$\delta^n_{v_i}$& Processing latency incurred by \gls{vnf} instance $v_i$ on node $n$	\\
		$\delta_t^{s,max}$ & Maximum latency bound permitted for traffic type $t$ \\
        $\zeta^n_{v_i}$& Total computational capacity consumed by \gls{vnf} instance $v_i$ on node $n$	\\
        $\theta^s_{v,t}$ & Binary indicator: $1$ if traffic type $t$ requires traversal of \gls{vnf} type $v$; 0 otherwise \\ \\
			\midrule[1.5pt] 
		\rowcolor{blue!7}
		\textbf{Decision Variable}& \textbf{Definition}\\		\midrule[1.5pt]
        $\phi_n$ & $1$ if physical node $n$ is powered on and hosting active \glspl{vnf}; $0$ otherwise\\
        $\beta^{n}_{v_i}$ & $1$ if \gls{vnf} instance $v_i$ is instantiated on physical node $n$; $0$ otherwise \\
        $\upgamma^{n,s}_{v_i,t}$ & $1$ if \gls{vnf} instance $v_i$ serving traffic type $t$  in slice $s$ is hosted on node $n$; $0$ otherwise  \\
		$\chi^{(n,m),s}_{(v_i,z_j),t}$ & $1$ if virtual link $(v_i,z_j)$ for traffic $t$ of slice $s$ is routed over physical link $(n,m)$; $0$ otherwise \\
        
        \bottomrule[1.25pt]
	\end{tabular}
\end{table}

\subsection{Architectural and Operational Assumptions}
To clearly scope the mathematical formulation and align it with realistic \gls{5g} deployments, the following architectural, computational, and economic assumptions are established.

\textbf{Multi-\gls{upf} Data Paths and \gls{3gpp} Compliance:} This model strictly considers the \gls{5g} data plane, exclusively orchestrating \glspl{upf}. The number of \glspl{upf} deployed along a single \gls{pdu} session data path is not constrained by \gls{3gpp} specifications, as detailed in Section 8.3.1 of \gls{3gpp} TS 23.501 \cite{3gpp.23.501} and formally acknowledged by the \gls{ietf} \cite{ietf_uplane}. To accurately model a realistic core network, we assume a multi-tiered \gls{upf} topology consisting of intermediate \glspl{upf} ($I-\gls{upf}$), Central \glspl{upf} ($C-\gls{upf}$), and \gls{pdu} Session Anchor \glspl{upf} ($P-\gls{upf}$). The $I-\gls{upf}$ acts as an edge branching node, routing latency-critical \gls{urllc} traffic directly to an edge \gls{dn} while forwarding \gls{embb} traffic deeper into the core. The $C-\gls{upf}$ acts purely as a transit node to mimic multi-hop core transit and does not connect to any \gls{dn}. Finally, the $P-\gls{upf}$ terminates the connection to the cloud \gls{dn}. Consequently, the \gls{pdu} data path for \gls{urllc} is modeled as ($\gls{gnb}  \rightarrow I-\gls{upf} \rightarrow \gls{dn}_{URLLC}$), whereas the path for \gls{embb} slices is modeled as ($\gls{gnb} \rightarrow I-\gls{upf} \rightarrow C-\gls{upf} \rightarrow P-\gls{upf} \rightarrow \gls{dn}_{eMBB}$). 

\textbf{Computational Overheads and Scaling:} To ensure compatibility with modern \gls{5g} testbed environments, \glspl{vnf} are assumed to be deployed as containerized \glspl{nf} that support dynamic scaling to adapt to varying traffic loads. For the purposes of this \gls{minlp} optimization, minor hardware-level execution delays are considered negligible. Specifically, the processing delay induced by load balancing across multi-core \gls{upf} instances is excluded. Furthermore, CPU context-switching delays, which scale linearly when shared physical cores alternate between processing traffic for different slices, are omitted, allowing the objective function to focus purely on total node capacity and link propagation delays. 

\textbf{Slice Isolation-as-a-Service Economics and Provisioning:} The model operates under the assumption that network slices (and their constituent user traffic) demand different tiers of security based on their operational criticality. In alignment with the Slice Isolation-as-a-Service framework, it is the responsibility of the network orchestration engine to strictly guarantee these requested isolation boundaries. We assume that the economic burden of provisioning isolated physical infrastructure, which inherently reduces multiplexing efficiency and utilizes more base hardware capacity, is offset by passing the extra costs to the tenants through premium service fees.

\subsection{Objective Function Formulation}

The primary goal of the optimization model is to jointly minimize the infrastructure resource footprint and the end-to-end latency of the user traffic, subject to the strict isolation bounds dictated by the Slice Isolation-as-a-Service framework. The objective function, formulated in Equation~(\ref{eq:obj-func}), consists of three distinct cost components.

The first term aggregates the total computational processing capacity, $\zeta^n_{v_i}$, consumed by all active \gls{vnf} instances across the network. The second term calculates the infrastructural baseline cost, where $P_n^B$ represents the provisioning overhead incurred when a physical node $n$ is powered on, which is tracked by the binary activation variable $\phi_n$. Finally, the third term accumulates the total end-to-end latency, $\delta^s_t$, experienced by all traffic types across every active network slice.

\begin{equation}
\min \left( 
    \sum_{v \in \mathcal{V}} \sum_{i \in I_v} \sum_{n \in \mathcal{N}} \tikzmarknode{left}{\zeta^n_{v_i}} 
    + \sum_{n \in \mathcal{N}} \tikzmarknode{middle}{P_n^B \cdot \phi_n} 
    + \sum_{s \in \mathcal{S}} \sum_{t \in \mathcal{T}_s} \tikzmarknode{right}{\delta^s_t} 
\right)
\begin{tikzpicture}[overlay,remember picture,red,>=stealth,shorten <=0.2ex,nodes={font=\small,align=left,inner ysep=1pt},<-]
    \path (left.south) ++ (0,-1.6em) node[anchor=north east] (diff1)
        {Total VNF\\ computational capacity};
    \draw (left.south) |- ([xshift=0.3ex]diff1.south west);
    
    \path (middle.south) ++ (0,-2em) node[anchor=north west] (diff2)
        {Active physical\\node overhead};
    \draw (middle.south) |- ([xshift=0.3ex]diff2.south east);
    
    \path (right.south) ++ (0.3,-0.2em) node[anchor=north west] (diff3)
        {End-to-end\\slice traffic delay};
    \draw (right.south) |- ([xshift=0.3ex]diff3.south east);
\end{tikzpicture}
\label{eq:obj-func}
\end{equation}
\vspace{3em}

\textbf{Subject to} constraints (\ref{eq:vnf-total-capacity}) to (\ref{eq:sec2}).

\subsubsection{\textbf{Computational Capacity and Traffic Latency}}
The \textbf{total processing capacity} $\zeta^n_{v_i}$ demanded by a specific \gls{vnf} instance $v_i$ hosted on physical node $n$ comprises two components: a static operational baseline and a dynamic traffic-dependent load. As expressed in Equation~(\ref{eq:vnf-total-capacity}), the baseline capacity $\zeta^{n, B}_{v_i}$ is allocated if the \gls{vnf} instance is instantiated (indicated by the binary variable $\beta^{n}_{v_i}=1$), while $\zeta^{n, T}_{v_i}$ represents the dynamic capacity required to process the active traffic load.

\begin{equation}
    \zeta^n_{v_i} = {\zeta^{n,B}_{v_i}} \cdot  {\beta^{n}_{v_i}} +  {\zeta^{n,T}_{v_i}} \quad \forall v \in \mathcal{V}, ~~ i \in I_v,~  n \in \mathcal{N}
    \label{eq:vnf-total-capacity}
\end{equation}

The dynamic traffic load, $\zeta^{n,T}_{v_i}$, scales proportionally with the number of slice traffic types routed through the instance. Assuming $\mu_v$ defines the unit processing capacity required by the \gls{vnf} type $v$, the total dynamic capacity is calculated in Equation~(\ref{eq:vnf-traffic-capacity}) by summing the requirements of all traffic types $t$ mapped to instance $v_i$ via the binary placement variable $\upgamma^{n,s}_{v_i,t}$.

\begin{equation}
    \zeta^{n,T}_{v_i}~=~
    \sum_{s \in \mathcal{S}}  \sum_{t \in \mathcal{T}_s}~
   {~\upgamma^{n,s}_{v_i,t} }\cdot{~\mu_{v}} \quad \forall v \in \mathcal{V}, ~~ i \in I_v,~  n \in \mathcal{N}
    \label{eq:vnf-traffic-capacity}
\end{equation}

The \textbf{total end-to-end latency} $\delta^s_t$ experienced by a traffic type $t$ is the sum of the nodal processing delays and the physical link propagation delays. In Equation~(\ref{eq:traffic-type-delay}), the first term aggregates the processing delay $\delta^n_{v_i}$ for every \gls{vnf} traversed by the traffic. The second term calculates the cumulative propagation delay by multiplexing the physical link delay $d(n,m)$  by the binary variable $\chi^{(n,m),s}_{(v_i,z_j),t}$, which equals 1 if the virtual link $(v_i,z_j)$ is mapped over the physical link $(n,m)$.

\begin{equation}
\begin{split}
\delta^s_t =  \sum_{v \in \mathcal{V}^s_t} \sum_{i \in I_v} \sum_{n \in \mathcal{N}} {\delta^n_{v_i}} \cdot \upgamma^{n,s}_{v_i,t} ~+~  \sum_{ (v_i, z_j) \in \mathcal{R}_t^s} \sum_{(n,m) \in \mathcal{L}}  {d(n,m)} \cdot \chi^{(n,m),s}_{(v_i,z_j),t} \quad \forall s \in \mathcal{S}, t \in \mathcal{T}_s
\end{split}
\label{eq:traffic-type-delay}
\end{equation}

Finally, the processing delay $\delta^n_{v_i}$ incurred at a specific \gls{vnf} instance is modeled as a queuing system. As defined in Equation~(\ref{eq:vnf-processing-delay}), it is a function of the \gls{vnf}'s average service rate $\omega_{v}$ and the total aggregate packet arrival rate $\lambda_t^s$ of all traffic types currently utilizing that specific instance.
\begin{equation}
\begin{split}
    \delta^n_{v_i} =  {1}/{(\omega_{v} - \sum_{s \in \mathcal{S}}  \sum_{t \in \mathcal{T}_s} \lambda_t^s \cdot \upgamma^{n,s}_{v_i,t})} \quad \forall s \in \mathcal{S},   t \in \mathcal{T}_s
\end{split}
\label{eq:vnf-processing-delay}
\end{equation}

The objective function is bounded by a rigorous set of constraints, which are divided into two logical categories: foundational infrastructural assignment constraints (detailed in Section~\ref{sec:assig_const}) and the novel Slice Isolation-as-a-Service constraints (detailed in Section~\ref{sec:siaas}).
  
\subsubsection{\textbf{Assignment constraints}} \label{sec:assig_const}

The assignment constraints govern the valid deployment of \gls{upf} instances and the mapping of virtual links onto the physical network. While these incorporate foundational \gls{vnf} placement principles commonly found in the literature \cite{Leyva-Pupo2022, Mei2020b}, they are specifically adapted here to operate at the granular traffic-type level to orchestrate data plane topologies.

First, Equation~(\ref{eq:domain-constraints}) restricts the decision variables to binary domains.

\begin{equation}
    \upgamma^{n,s}_{v_i,t}~, ~\chi^{(n,m),s}_{(v_i,z_j),t}~, ~ \beta^{n}_{v_i}~,~\phi_n ~ \in \{0,1\} 
\label{eq:domain-constraints}
\end{equation}

Equation~(\ref{eq:all-vnf-deployed}) guarantees mapping completeness: if a traffic type $t$ belonging to slice $s$ requires a specific \gls{vnf} type $v$ (indicated by parameter $\theta^s_{v,t} = 1$), it must be mapped to exactly one active instance $v_i$ across the physical infrastructure.

\begin{equation}
    \sum_{  n \in \mathcal{N}} \sum_{i \in I_v}   \upgamma^{n,s}_{v_i,t} = \theta^s_{v,t} \quad \forall s \in \mathcal{S}, ~ t \in \mathcal{T}_s, v \in \mathcal{V}
    \label{eq:all-vnf-deployed}
\end{equation}

To prevent resource wastage, Equations~(\ref{eq:vnf-launch-constrnt1}) and (\ref{eq:vnf-launch-constrnt2}) utilize the Big-M method to ensure that a \gls{vnf} instance is instantiated ($\beta^{n}_{v_i} = 1$) if and only if at least one traffic type is actively routed through it. Here, $\mathcal{M}$ is an arbitrarily large constant exceeding the maximum possible number of traffic types mapped to an instance.

\begin{equation}
    \sum_{  s \in \mathcal{S}} \sum_{t \in \mathcal{T}_s}   \upgamma^{n,s}_{v_i,t} \le \mathcal{M} \cdot \beta^{n}_{v_i} , \quad \forall v \in \mathcal{V}, ~ i \in I_v, ~ n \in \mathcal{N}
    \label{eq:vnf-launch-constrnt1}
\end{equation}

\begin{equation}
     \beta^{n}_{v_i} - 	\sum_{  s \in \mathcal{S}} \sum_{t \in \mathcal{T}_s}   \upgamma^{n,s}_{v_i,t} \le 0, \quad \forall v \in \mathcal{V}, ~ i \in I_v, ~ n \in \mathcal{N}
    \label{eq:vnf-launch-constrnt2}
\end{equation}

Equation~(\ref{eq:vnf-instance-constrnt}) restricts each unique \gls{vnf} instance $i$ of type $v$ to be deployed on at most a single physical node $n$.

\begin{equation}
    \sum_{i \in I_v} \beta^{n}_{v_i} \le 1, \quad \forall v \in \mathcal{V}, ~n \in \mathcal{N} 
    \label{eq:vnf-instance-constrnt}
\end{equation}
				
Computational resources are strictly bounded at both the virtual and physical levels. Equation~(\ref{eq:vnf-capcty-limt}) ensures the total dynamic capacity demanded by an active \gls{vnf} instance does not exceed its defined hardware limit $\zeta^{max}_{v}$.

\begin{equation}
    \zeta^n_{v_i} \le  {\zeta^{max}_{v}}, \quad \forall n \in \mathcal{N},~~  v \in \mathcal{V}, ~~ i \in I_v
    \label{eq:vnf-capcty-limt}
\end{equation}

Furthermore, Equation~(\ref{eq:node-capcty-limt}) ensures that the cumulative capacity of all \gls{vnf} instances deployed on a given physical node $n$, combined with its baseline operating capacity $P_n^B$, does not exceed the absolute capacity limit of the node $C^{max}_{n}$.

\begin{equation}
    \sum_{v \in \mathcal{V}} \sum_{i \in I_v} \zeta^n_{v_i}  + P_n^B \cdot \phi_n ~~\le ~ C^{max}_{n}, \quad \forall n \in \mathcal{N} 
    \label{eq:node-capcty-limt}
\end{equation}

Equation~(\ref{eq:node-activation}) governs physical node activation, dictating that a node's operational status indicator $\phi_n$ must be set to 1 if any active \gls{vnf} instance is hosted upon it.

\begin{equation}
    \beta^{n}_{v_i} \leq \phi_n \quad \forall v \in \mathcal{V}, ~~ i \in I_v,~  n \in \mathcal{N}
    \label{eq:node-activation}
\end{equation}

For data path routing, Equation~(\ref{eq:phy-to-vir-links-mapping}) defines the non-linear relationship linking virtual and physical topologies. It states that a physical link $(n,m)$ can only host the virtual link $(v_i, z_j)$ if the communicating \gls{vnf} instances $v_i$ and $z_j$ are physically mapped to nodes $n$ and $m$, respectively.

\begin{equation}
    \begin{split}
        \chi^{(n,m),s}_{(v_i,z_j),t} \le \upgamma^{n,s}_{v_i,t} ~ .~ \upgamma^{m,s}_{z_j,t}  \quad\quad \forall   (m, n) \in \mathcal{L}, ~ (v_i, z_j) \in \mathcal{R}, ~ t \in \mathcal{T}
    \end{split}
    \label{eq:phy-to-vir-links-mapping}
\end{equation}

Equation~(\ref{eq:link-capcity-constrnt}) bounds the network traffic, ensuring the cumulative bandwidth required by all traffic types traversing physical link $(n,m)$ remains within the link's maximum capacity $b(n,m)$.
\begin{equation}
    \begin{split}
        \sum_{s \in \mathcal{S}}  \sum_{t \in \mathcal{T}_s} \sum_{ (v_i, z_j) \in \mathcal{R}_t^s}  \lambda_t^s \cdot  \chi^{(n,m),s}_{(v_i,z_j),t} ~\le ~ b(n,m)  \quad\quad \forall (n,m) \in \mathcal{L}
    \end{split}
    \label{eq:link-capcity-constrnt}
\end{equation}

Finally, Equation~(\ref{eq:traffic-type-delay-constrnt}) enforces the \gls{sla} delay bounds, guaranteeing that the calculated end-to-end latency $\delta^s_t$ for any traffic type does not exceed its maximum tolerated threshold $\delta^{s,max}_t$.

\begin{equation}
    \delta^s_t \le \delta^{s,max}_t ~ ~~~~~\forall   s \in \mathcal{S}, ~ t \in \mathcal{T}_s
\label{eq:traffic-type-delay-constrnt}
\end{equation}

\subsubsection{\textbf{Slice Isolation-as-a-Service Security Constraint}} \label{sec:siaas}
The fundamental novelty of the Slice Isolation-as-a-Service framework is its ability to proactively defend against cross-slice attacks, such as the Reflection attack and \gls{dp} Flooding evaluated in Section~\ref{sec:results_gtp} and \ref{sec:results_flood}, by enforcing strict isolation boundaries based on requested security tiers.

\begin{enumerate}[leftmargin=*]
    \item \textbf{High-Security Slice:}
    Slices requiring a `high' security level necessitate strict physical isolation to prevent hardware-level resource contention and cross-slice vulnerabilities. Constraint~(\ref{eq:sec1}) ensures that if the required security level of a slice ($s_h$) is `high', the virtual and physical resources used by this slice are strictly isolated and not shared with any other slices.

    \begin{equation}
    \begin{split}
        \tikzmarknode{sec1_left}{\upgamma^{n_h,s_h}_{v^h_i,t_h}} = 1 \implies
        \sum_{ s \in \mathcal{S}} \sum_{t \in \mathcal{T}_s} \sum_{v \in \mathcal{V}} \sum_{i \in I_v} \tikzmarknode{sec1_right}{\upgamma^{n_h,s}_{v_i,t}} \le 1 , \quad ~~~~~ \quad \forall n_h \in \mathcal{N}
    \end{split}
    \begin{tikzpicture}[overlay,remember picture,red,>=stealth,shorten <=0.2ex,nodes={font=\small,align=left,inner ysep=1pt},<-]
        \path (sec1_left.south) ++ (0,-0.8em) node[anchor=north east] (diff_s1l)
            {If high-security slice $s_h$ \\ uses physical node $n_h$};
        \draw (sec1_left.south) |- ([xshift=0.3ex]diff_s1l.south west);
        
        \path (sec1_right.south) ++ (0,-0.8em) node[anchor=north west] (diff_s1r)
            {Physical node $n_h$ is not \\ shared with other slices};
        \draw (sec1_right.south) |- ([xshift=0.3ex]diff_s1r.south east);
    \end{tikzpicture}
    \label{eq:sec1}
\end{equation}
\vspace{1.5em}

    Where $s_h$ represents a slice requesting high security, while $t_h$, $v_i^h$, and $n_h$ represent the specific traffic type, \gls{vnf} instance, and physical node used by slice $s_h$, respectively.

    \item \textbf{Medium-Security Slice:}
    Slices requiring a `medium' security level necessitate logical isolation when utilizing virtual resources. Constraint~(\ref{eq:sec2}) ensures that if a slice ($s_m$) requires a medium security level, it is isolated at the virtual resource level. Specifically, the \gls{vnf} instances assigned to it are strictly dedicated and not shared with other slices.

    \begin{equation}
    \begin{split}
        \tikzmarknode{sec2_left}{\upgamma^{n_m,s_m}_{v^m_i,t_m}} = 1 \implies
        \sum_{s \in \mathcal{S}} \sum_{t \in \mathcal{T}_s}
        \tikzmarknode{sec2_right}{\upgamma^{n_m,s}_{v^m_i,t}} \le 1 , ~~~~~ \quad \quad \forall v^m \in \mathcal{V}, ~ i \in I_v, ~ n_m \in \mathcal{N}
    \end{split}
    \begin{tikzpicture}[overlay,remember picture,red,>=stealth,shorten <=0.2ex,nodes={font=\small,align=left,inner ysep=1pt},<-]
        \path (sec2_left.south) ++ (0,-0.8em) node[anchor=north east] (diff_s2l)
            {If medium-security slice $s_m$ \\ uses \gls{vnf} instance $v^m_i$};
        \draw (sec2_left.south) |- ([xshift=0.3ex]diff_s2l.south west);
        
        \path (sec2_right.south) ++ (0,-0.8em) node[anchor=north west] (diff_s2r)
            {\gls{vnf} instance $v^m_i$ is not \\ shared with other slices};
        \draw (sec2_right.south) |- ([xshift=0.3ex]diff_s2r.south east);
    \end{tikzpicture}
    \label{eq:sec2}
\end{equation}
\vspace{2.5em}

    Where $s_m$ represents a slice requesting medium security, while $t_m$, $v_i^m$, and $n_m$ represent the specific traffic type, \gls{vnf} instance, and physical node used by slice $s_m$, respectively.
    
    \item \textbf{Low-Security Slice:} A slice requiring low slice security has no restrictions on the isolation of resources. To maximize multiplexing efficiency and minimize operational costs, the \gls{minlp} solver is permitted to freely share both the \gls{vnf} instances and the physical nodes among low-security slices, bounded only by the standard capacity constraints defined in Section~\ref{sec:assig_const}. Table~\ref{tab:Security_levels} summarizes the level of isolation provided for the different security levels.  
\end{enumerate}

\section{Design and Development of a Containerized Multi-UPF 5G Security Testbed}\label{sec:testbed}
This section details the \gls{5g} testbed setup, the underlying emulation tools, and the engineering challenges overcome during its deployment.

\subsection{5G Emulation Infrastructure}
\begin{figure}[t]
    \centering
    \includegraphics[width=0.75\linewidth]{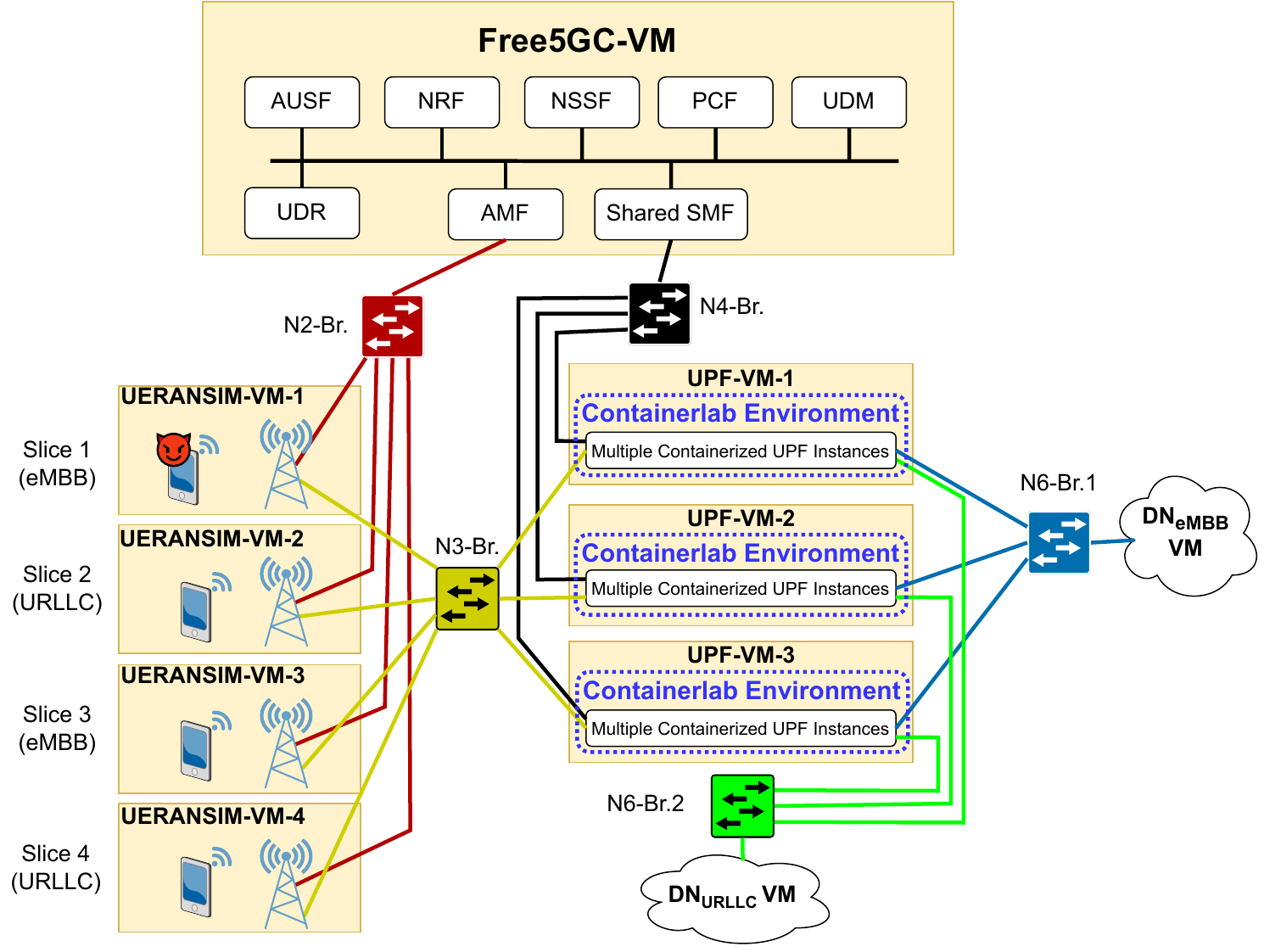}
    \caption{5G testbed architecture deployed across a multi-VM environment using Free5GC and containerized UPFs}
    \label{fig:testbed}
\end{figure}

We developed a private \gls{5g} testbed at Carleton University utilizing open-source tools. The \gls{5g} testbed is implemented on a server with a 32-core Intel Xeon processor and 64 GB of RAM. Oracle VirtualBox \cite{virtualbox} \glspl{vm} are used to deploy different parts of the \gls{5g} network, including the control plane, data plane, \gls{ran}, and \gls{dn}. The testbed comprises 10 VirtualBox \glspl{vm}, with one \gls{vm} dedicated to running the \gls{5g} core using Free5GC, three \glspl{vm} to run the data plane \glspl{upf}, four \glspl{vm} to emulate \gls{ran} using UERANSIM, and two \glspl{vm} representing \glspl{dn} for user traffic. The testbed is shown in Figure~\ref{fig:testbed} 

\textbf{\gls{5g} Core} is deployed using the Linux Foundation's open-source Free5GC project \cite{free5gc}, utilizing \texttt{go-upf} (version 1.2.7)~\cite{go_upf} and the \texttt{gtp5g} Linux kernel module (version 0.9.15)~\cite{gtp5g} for the \gls{dp}. Free5GC supports essential \gls{5g} core \glspl{nf} including \gls{amf}, \gls{smf}, \gls{upf}, and other \gls{cp} \glspl{nf}. A dedicated \gls{vm} runs \gls{5g} \gls{cp} \glspl{nf}, while three \glspl{vm} are used to host \gls{dp} \glspl{upf} to forward \gls{ue} traffic between \gls{ran} and \gls{dn}. As mentioned in Section~\ref{sec:Model}, we opted for a multi-\gls{upf} data plane architecture to imitate an actual network and to realize an edge and cloud network scenario using multiple \glspl{dn}. Each \gls{upf} \gls{vm} contains 3 types of \glspl{upf}, including $I-\gls{upf}$, $C-\gls{upf}$, and $P-\gls{upf}$, and each type can have a maximum of 3 instances in a single \gls{vm}, as set in the parameters of the optimization model. To deploy and interconnect the \glspl{upf}, Nokia Containerlab \cite{containerlab} is used, and each \gls{upf} instance runs in dedicated containers. Further details are explained in Section~\ref{sec:containerlab}.

\textbf{\gls{5g} \gls{ran}} is implemented with UERANSIM, an open-source emulator to run \gls{ue} and \gls{gnb}. Four \glspl{vm} are assigned to initiate four \gls{ran} instances, each running a \gls{ue} and a \gls{gnb}. Each \gls{ran} instance is assigned to a specific slice in the network. The UERANSIM integrates seamlessly with the core, Free5GC, and allows the \glspl{ue} to connect with their respective \glspl{dn}. Two dedicated \glspl{vm} serve as $\gls{dn}_{URLLC}$ and $\gls{dn}_{eMBB}$, each serving \gls{urllc} and \gls{embb} traffic, respectively. Utilizing private \gls{vm}-based \glspl{dn} enables the testing of attacks and security measures without the need to connect to the public internet. All \glspl{vm} connect and communicate using VirtualBox's Internal Network \cite{vbox_inet}, and the \gls{5g} network interfaces have their own isolated internal network. For example, \gls{ran} and \gls{5g} core \glspl{vm} connect using the N2 internal network, while the \gls{ran} and \gls{upf} \glspl{vm} connect using the N3 internal network, the \gls{upf} and \gls{5g} core (\gls{smf}) \glspl{vm} connect using the N4 internal network, and the \gls{upf} and \gls{dn} \glspl{vm} connect using the N6 internal network. These testbed and networking settings ensure that the traffic is separated and the experimentation is performed in a controlled environment without connecting to the public internet.

\subsection{Containerized Data Plane Deployment via Containerlab} \label{sec:containerlab}
To strictly replicate the complex, optimization-driven resource allocations calculated in Section~\ref{sec:Model}, the testbed required a highly flexible data plane capable of hosting multiple distinct \gls{upf} instances per node. To the best of our knowledge, this study presents the first \gls{5g} testbed implementation to leverage Nokia Containerlab \cite{containerlab} to orchestrate a distributed, multi-interface \gls{upf} data plane. 

Developing this custom architecture was a significant engineering contribution motivated by the limitations of standard \gls{5g} deployments. Most existing open-source \gls{5g} academic testbeds utilize single-machine deployments, which lack the multi-node flexibility required to realistically mimic distributed edge-to-cloud architectures. In contrast, attempting to run multiple Free5GC \glspl{upf} as standard native Linux processes on a single \gls{vm} results in fatal kernel module conflicts (detailed further in Section~\ref{sec:testbed_challenges}).

To achieve multi-\gls{upf} instantiation while preserving the realistic multi-node \gls{vm} topology, the \glspl{upf} were containerized. However, a \gls{3gpp}-compliant \gls{5g} data plane requires complex network wiring; each \gls{upf} must simultaneously maintain distinct, isolated interfaces, specifically N3 (\gls{ran}), N4 (Control Plane), N6 (\gls{dn}), and N9 (inter-\gls{upf}), to communicate with various network segments. Standard Docker environments heavily abstract this networking, making external routing highly restrictive.

Containerlab was strategically adopted because it natively solves this multi-interface wiring problem. It provided the granular network namespace control required to successfully wire the containerized \glspl{upf} to the external \gls{5g} core, \gls{ran}, and \gls{dn} components hosted on completely separate \glspl{vm}. Furthermore, Containerlab enabled the automated, repeatable deployment of these highly customized topologies through a single declarative YAML file, injecting IP assignments and configuration files via \texttt{exec} blocks upon boot. This novel approach successfully delivered the advanced, multi-interface routing capabilities of a Kubernetes environment (which typically requires complex Helm charts and the Multus CNI plugin) without forcing an overly complex overhaul of the lightweight, \gls{vm}-based core infrastructure.

\subsection{Implementation Challenges and Resolution} \label{sec:testbed_challenges}
Developing a multi-node, virtualized \gls{5g} testbed presented several challenges, specifically regarding \gls{dp} \gls{upf} deployment, network connections, and high-rate traffic generation.

\textbf{Multi-\acrshort{upf} Instantiation and Namespace Conflicts:} As the optimization model in Section~\ref{sec:Model} considers deploying multi-\gls{upf} instances on a single physical server, developing a \gls{5g} testbed with multiple \gls{upf} instances in one \gls{vm} was the primary hurdle. Initial attempts to run multiple Free5GC \gls{upf} instances as standard Linux processes failed because Free5GC relies on the \texttt{gtp5g} kernel module \cite{gtp5g}, and trying to bind multiple native processes to this module in the same network namespace resulted in a fatal interface conflict error, as reported in \cite{gtp5g_error}. To resolve this namespace conflict, Docker containers \cite{docker} were used along with Nokia Containerlab \cite{containerlab} to run and isolate multiple \gls{upf} instances, as detailed in Section~\ref{sec:containerlab}.

\textbf{Container-to-VM Network Bridging:} Transitioning to a containerized \gls{upf} introduced a secondary challenge of bridging the isolated container namespaces to communicate with the external \glspl{nf} (\gls{smf}) and network components (\gls{ran} and \gls{dn}) running on separate and dedicated \glspl{vm}. By default, containers are isolated behind a local bridge, making external routing complex and often reliant on performance-degrading NAT rules.

To ensure Layer-2 reachability across the testbed, MACVLAN interfaces \cite{clab_macvlan} were utilized on the \gls{upf} containers to connect to N3, N4, and N6 interfaces of the host \gls{vm}. This effectively assigned them \gls{ip} addresses on the same subnets as the external \glspl{vm} and allowed seamless communication. On the other hand, the N9 interface, required for local \gls{upf} communication within the same \gls{vm}, was implemented using a standard bridge.

\textbf{\acrshort{ran} Emulation Instability Under Flooding Attack:} During high-volume flooding attack generation, the attacking \gls{ue}'s radio link becomes unstable, triggering frequent crashes and halting the injection of flood traffic into the \gls{upf}, as reported in \cite{uransim_error}. To overcome this emulation bottleneck and ensure a sustained, high-rate volumetric flood traffic, the attack generation was decoupled from UERANSIM. Flood attack traffic was first successfully transmitted through the \gls{gnb} and captured as a \gls{pcap} file containing network-valid, session-specific \gls{gtp}-U headers. The TRex traffic generator \cite{trex} was then deployed to ingest this \gls{pcap} file and replay the encapsulated packets directly into the N3 interface at a rate of 10,000 packets per second, effectively bypassing the susceptible \gls{ran} emulation while maintaining protocol compliance.

\section{Experimental Topologies and Methodology} \label{sec:exp_topology}
This section details the network configurations obtained from the optimization model and implemented in our \gls{5g} testbed. Additionally, we explain the evaluation methodology, the user traffic tested, and the metrics used to evaluate the performance of the network slices.

\begin{table}[tbp]
    \centering
	\captionof{table}{Simulation Parameters (a) to Obtain Testbed Topologies and (b) for Performance Evaluation of the Optimization Model. } \label{tab:sim_paramter}
	\begin{tabular}{l||c}\toprule[1.5pt]
        \rowcolor{gray!20} \multicolumn{2}{c}{(a) Simulation Parameters to Obtain Testbed Topologies}\\\midrule[1.5pt]
		\rowcolor{blue!7}
		\bf Parameter		 & \bf Value 	\\\midrule[1.5pt]
		Number of physical nodes          &  3 nodes			\\
		\rowcolor{blue!7}
		Maximum capacity of physical nodes        &	2.7 GHz	\\
        Base provisioning capacity of the physical node & 135 MHz \\ \rowcolor{blue!7}
		Physical link delay              &	2 $ms$		     \\
		Physical link maximum bandwidth   &	35 Mbps			  \\
        \rowcolor{blue!7}
		Maximum capacity of \gls{vnf} instance   &  2.7 GHz	   \\
		Number of instances per \gls{vnf}    &  3 instances	        \\
        \rowcolor{blue!7}
		\glspl{vnf} base capacity        &  10 MHz	\\
		\glspl{vnf} capacity required to serve traffic    &  27 - 135 MHz	\\
        \rowcolor{blue!7}
        \glspl{vnf} service rate         &  1000 - 2000 packets/sec	      \\
        Traffic packet rate for all slices    &  50 packets/sec	\\ \rowcolor{blue!7}
		Average packet size & 	1 KB 		\\
        Number of network slices & 4 slices \\\midrule[1.5pt]
        \rowcolor{gray!20}  \multicolumn{2}{c}{(b) Simulation Parameters for Performance Evaluation}\\\midrule[1.5pt]
        \rowcolor{blue!7}
        \bf Parameter		 & \bf Value 	\\\midrule[1.5pt]
        Number of physical nodes          &  5 nodes			\\ \rowcolor{blue!7}
        Number of network slices & 2 - 10 slices \\
		\bottomrule[1.25pt]
		\end {tabular}\par
\end{table}

\subsection{Optimization-Driven Slice Allocation}
\par To transition from the theoretical optimization model to testbed implementation, we run the model using the simulation parameters in Table~\ref{tab:sim_paramter}(a). The parameter values are obtained through a hybrid approach: physical node and \gls{vnf} maximum capacities ($2.7~GHz$ capacity and $3$ instances per \gls{vnf}) and physical link characteristics ($2~ms$ delay and $35~Mbps$ bandwidth) were obtained empirically from our testbed, while \glspl{vnf}' service rates and capacities required to serve traffic were adopted from existing telecommunication literature \cite{sattar2025secure, Leyva-Pupo2022, souza2024enabling}.

Table~\ref{tab:opt_exp_conf} shows seven experimental scenarios, each outlining distinct security isolation configurations (Low, Medium, High) of the four network slices. We simulate a combination of various security levels for each slice to obtain multiple resource allocation assignments. To practically evaluate the performance impacts of these isolation strategies, we select three specific topologies for deployment in the testbed, corresponding to Experiments 1, 3, and 5. These three topologies were chosen because they encompass all primary security isolation levels requested by the \gls{embb} (Slices 1 and 3) and \gls{urllc} (Slices 2 and 4) traffic profiles.

\begin{table*}[t!]
    \centering
\caption{Requested Slice Security Levels for Optimization Model. The \textcolor{yellow!100}{\textbf{Yellow}} colored columns represent the experimental topologies implemented in our testbed.}
\label{tab:opt_exp_conf}
    \begin{tabular}{l||l|l|l|l|l|l|l}\toprule[1.5pt]
        \rowcolor{blue!7}
         &  \multicolumn{7}{c}{\textbf{Requested Slice Security Levels}}\\ \cline{2-8}
         \rowcolor{blue!7} 
         \bf Slice Types &  \cellcolor{yellow!80} \bf Exp 1& \bf Exp 2&  \cellcolor{yellow!80} \bf Exp 3& \bf Exp 4&  \cellcolor{yellow!80} \bf Exp 5& \bf Exp 6&  \bf Exp 7\\\hline
         Slice 1 - eMBB&  \cellcolor{yellow!80}Low&  Low&  \cellcolor{yellow!80}Low&  Low&  \cellcolor{yellow!80}Low&  Low& Low\\\hline
         \rowcolor{blue!7}
         Slice 2 - URLLC&  \cellcolor{yellow!80}Low&  High&  \cellcolor{yellow!80}High&  Medium&  \cellcolor{yellow!80}Medium&  High& Medium\\\hline
         Slice 3 - eMBB&  \cellcolor{yellow!80}Low&  Low&  \cellcolor{yellow!80}Low&  Low&  \cellcolor{yellow!80}Low&  Low& Low\\\hline
         \rowcolor{blue!7}
         Slice 4 - URLLC&  \cellcolor{yellow!80}Low&  Low&  \cellcolor{yellow!80}High&  Low&  \cellcolor{yellow!80}Medium&  Medium& High\\ \bottomrule[1.25pt]
    \end{tabular}
\end{table*}

\begin{figure}[bthp]
    \centering
    \begin{subfigure}[b]{0.32\textwidth}
        \centering
        \includegraphics[width=\textwidth]{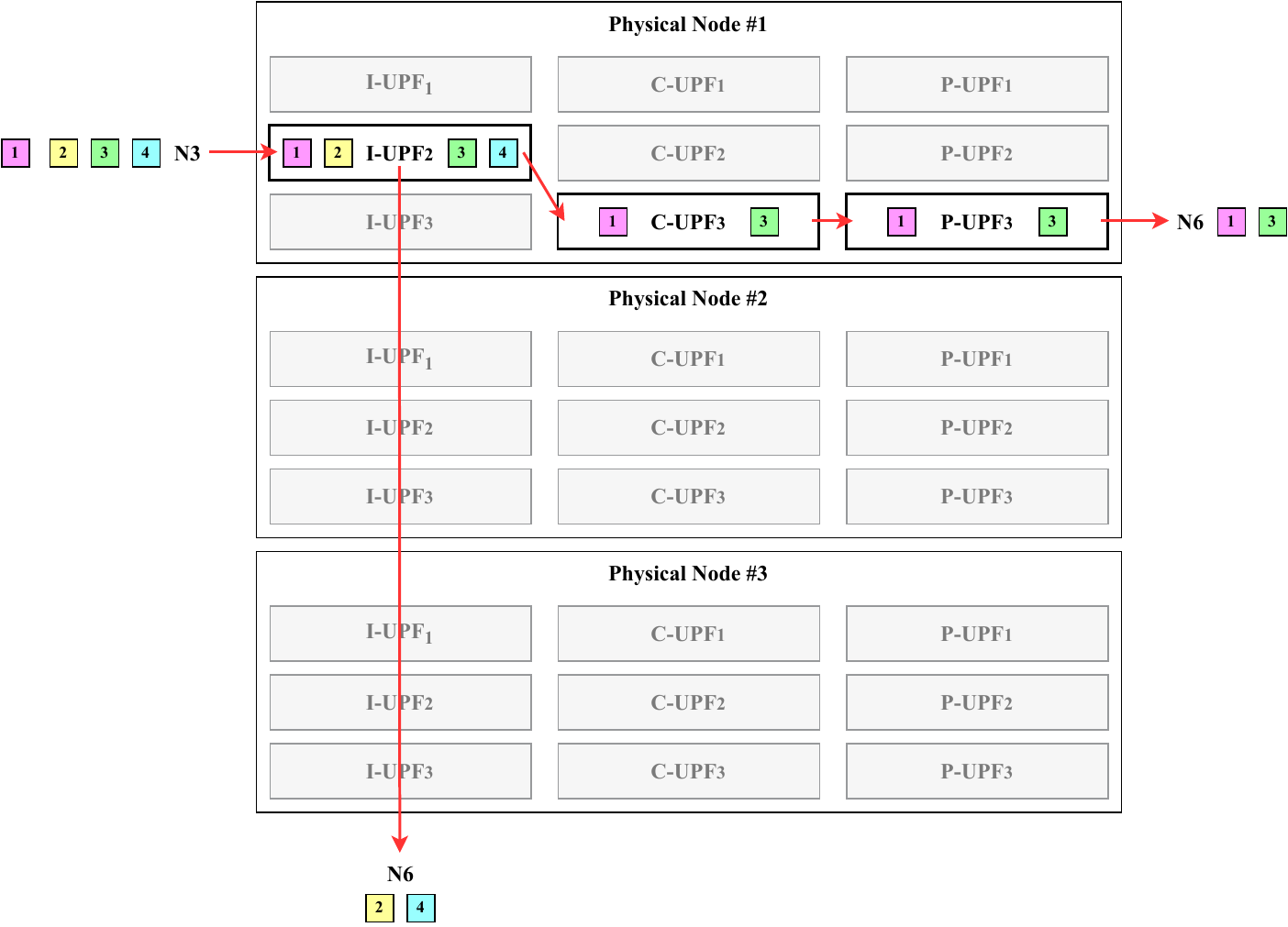}
        \caption{} 
        \label{fig:topology1}
    \end{subfigure}
    \hfill 
    \begin{subfigure}[b]{0.32\textwidth}
        \centering
        \includegraphics[width=\textwidth]{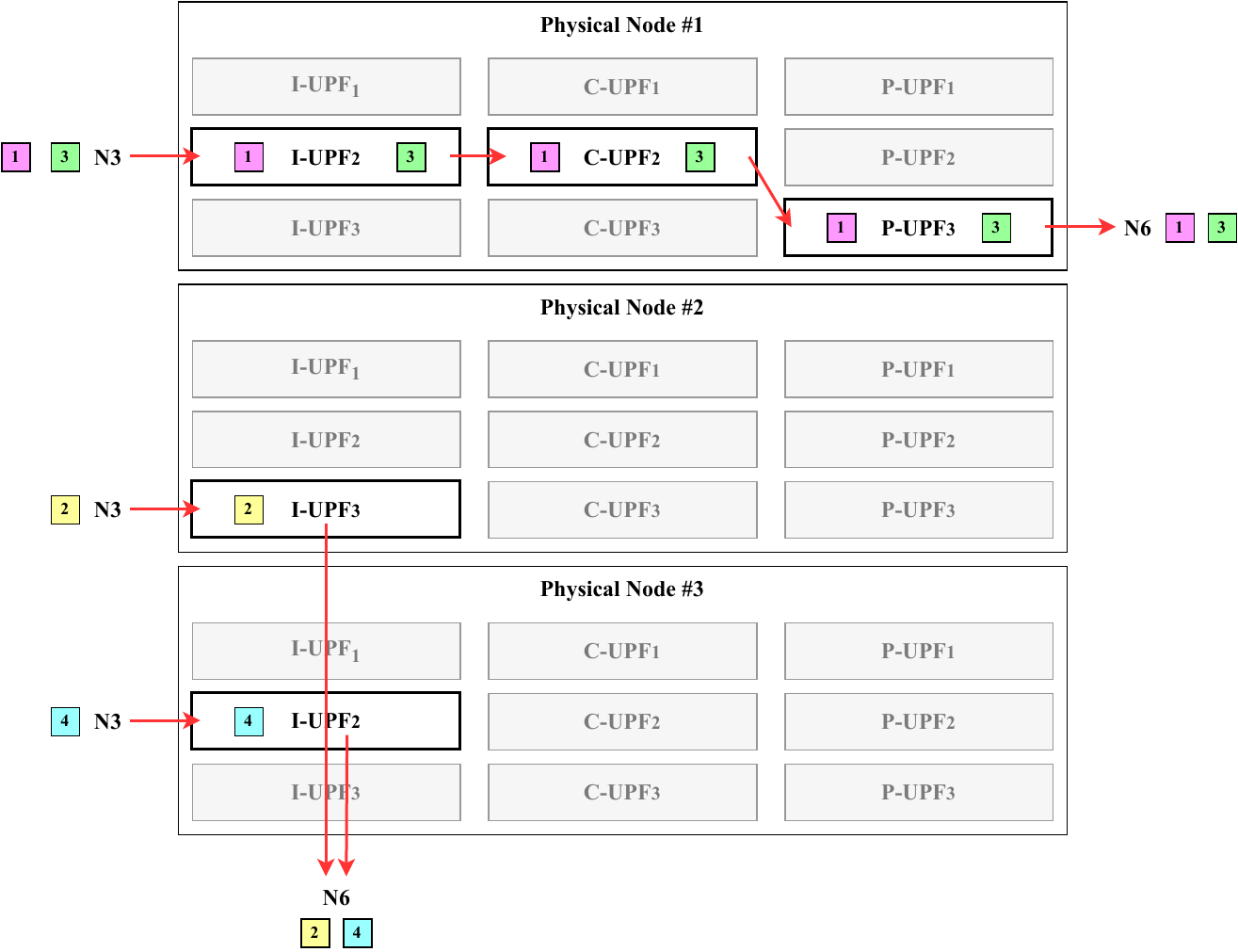}
        \caption{} 
        \label{fig:topology2}
    \end{subfigure}
    \hfill
    \begin{subfigure}[b]{0.32\textwidth}
        \centering
        \includegraphics[width=\textwidth]{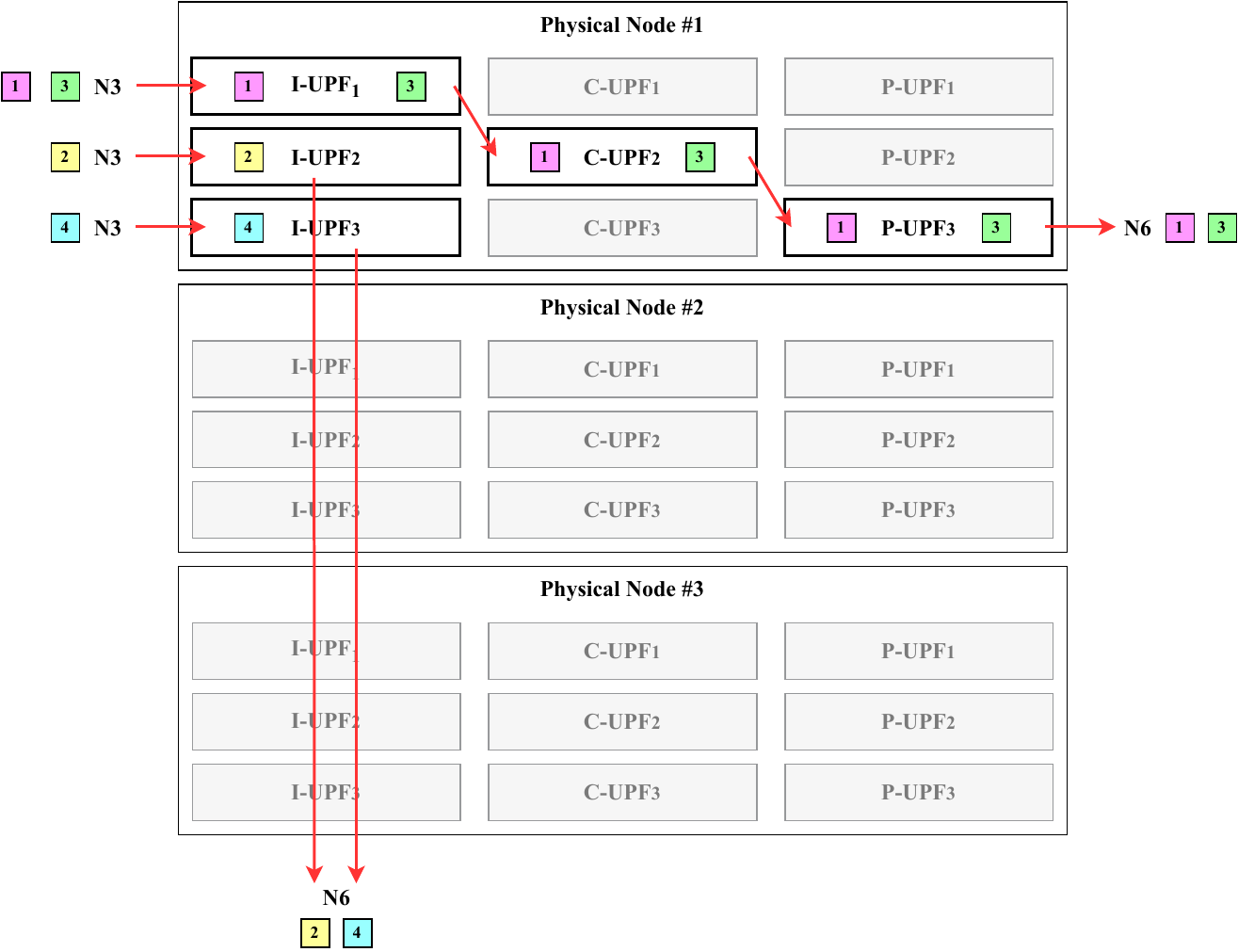}
        \caption{} 
        \label{fig:topology3}
    \end{subfigure}
    
    \vspace{1em} 
    \includegraphics[width=0.5\textwidth]{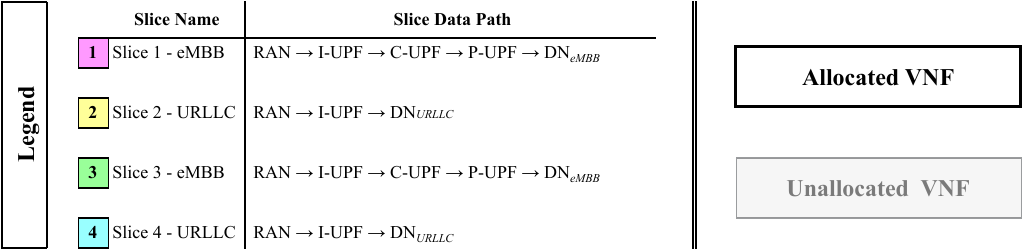}
    \caption{Network Topologies Obtained from the Optimization Model. (a) Allocated Resources for Experiment \#1 ( Topology \#1). (b) Allocated Resources for Experiment \#3 (Topology \#2). (c) Allocated Resources for Experiment \#5 (Topology \#3). The legend at the bottom applies to all three sub-figures.}
    \label{fig:Topologies}
    
\end{figure}

The resulting resource allocations, driven by the optimization model, are illustrated in Figures~\ref{fig:topology1} to \ref{fig:topology3}.

\begin{itemize}
    \item \textbf{Topology 1 (Experiment 1 - Full Sharing):} All four slices requested a ``Low'' security level; hence, the optimization model maximizes resource efficiency by assigning all slices to share a single $I-\gls{upf}_2$, as shown in Figure~\ref{fig:topology1}. Additionally, \gls{embb} slices (Slice 1 and Slice 3) share the $C-\gls{upf}_3$ and $P-\gls{upf}_3$ to reach their \gls{dn}. All four slices share the underlying physical and virtual resources.
    \item \textbf{Topology 2 (Experiment 3 - Strict Isolation):} The \gls{urllc} slices (Slice 2 and Slice 4) requested a ``High'' security level, while the \gls{embb} slices (Slice 1 and Slice 3) requested ``Low''. Figure~\ref{fig:topology2} demonstrates strict isolation, where Slice 2 is allocated a dedicated and isolated $I-\gls{upf}_3$ deployed in physical node 2. Similarly, Slice 4 is allocated an isolated $I-\gls{upf}_2$ running on physical node 3. Slices 1 and 3 remain aggregated on $I-\gls{upf}_2$, $C-\gls{upf}_2$, and $P-\gls{upf}_3$ deployed in physical node 1.
    \item \textbf{Topology 3 (Experiment 5 - Medium Isolation):} The \gls{urllc} slices requested a ``Medium'' security level. Figure~\ref{fig:topology3} illustrates a hybrid resource assignment where Slice 2 and Slice 4 are assigned to isolated virtual resources $I-\gls{upf}_2$ and $I-\gls{upf}_3$, respectively, but both slices share the same physical node along with Slices 1 and 3. \gls{embb} slices are also sharing the $I-\gls{upf}_1$, $C-\gls{upf}_2$, and $P-\gls{upf}_3$ in physical node 1.
\end{itemize}

\subsection{Evaluation Methodology} \label{sec:Eval_Method}
To comprehensively evaluate the performance and resilience of the proposed optimization model, the implemented topologies are assessed through three distinct experiments in the presence of the attacks described in Section~\ref{sec:methodology}. These experiments measure reflection attack success rate, latency degradation, packet loss rate, and bandwidth starvation for \glspl{ue} across different slice isolation levels.

\subsubsection{\textbf{Traffic  Injection using Reflection Attack}} \label{sec:reflection}
To analyze the susceptibility of \gls{dp} \glspl{upf} to known attacks, all three topologies were subjected to the \gls{gtp}-in-\gls{gtp} reflection attack, described in Section~\ref{sub:attacks}. In this experiment, Slice 1 and Slice 2 were configured as the attacking slices, while Slice 3 and Slice 4 were designated as the target victim slices. Each attacking slice targets both victim slices, and we measure the \textbf{attack success rate} across different slice isolation configurations. Attack success rate is a binary metric that determines whether the spoofed reflection traffic successfully breached the isolation boundaries to reach the victim \gls{ue}. We also measure the \textbf{cumulative traffic injected} into the victim \gls{ue}'s tunnel, which quantifies the severity of the victim's data quota exhaustion and billing fraud.

\subsubsection{\textbf{\gls{udp} Probing during Flooding Attack}}
To evaluate the impact of a TRex-driven flooding attack on slice latency and stability, constant-rate \gls{udp} traffic was generated using a custom Python script. Legitimate \glspl{ue} transmitted \gls{udp} probes to their \glspl{dn} at a steady rate of 20 probes per second, simulating latency-sensitive \gls{urllc} telemetry or \gls{embb} live-streaming applications. The performance of legitimate slices under a flooding attack was evaluated using the following metrics:

\begin{itemize}
    \item \textbf{\gls{rtt}:} For every 20 probes per second, the maximum, minimum, and average \gls{rtt} values were captured to determine the degradation in latency. To ensure absolute precision against virtualization clock synchronization anomalies, Python's \texttt{time.perf\_counter()} was utilized for \gls{rtt} calculations.
    \item \textbf{Jitter:} The average jitter was calculated to assess the variability and instability in packet \gls{rtt} caused by resource contention at the \gls{upf}.
    \item \textbf{Packet Loss:} The average percentage of lost \gls{udp} probes during the attack was recorded, including the $95\%$ confidence intervals to ensure statistical reliability of the slice isolation comparisons across the three topologies.
\end{itemize}

\subsubsection{\textbf{\gls{tcp} Throughput during Flooding Attack}}
To examine the flooding attack's impact on legitimate \glspl{ue}' data transfer capabilities (e.g., simulating an \gls{embb} large file download or buffered video stream), \gls{tcp} traffic streams were generated using the iPerf3 network testing tool. The primary metric extracted from this experiment was the \gls{tcp} throughput observed using iPerf3. By capturing the throughput before, during, and after the flooding attack window, this metric directly quantifies the available bandwidth capacity and resource starvation experienced by legitimate \glspl{ue}.

\section{Insights from Reflection Attack Experiments} \label{sec:results_gtp}
This section analyzes the empirical results gathered from the \gls{5g} testbed across the three optimization-driven topologies. By evaluating these distinct resource allocation strategies, we quantify the architectural resilience of \gls{5g}/\gls{6g} network slices against attacks. 

A recent ACM \gls{ccs} 2025 study \cite{shaik2025uncovering} has demonstrated the mechanics of the \gls{gtp}-in-\gls{gtp} protocol tunneling vulnerability; however, this study only evaluated unsliced core deployment scenarios. An implicit assumption in \gls{5g} architecture is that network slicing inherently provides security isolation \cite{kotulski2017end}. The results of our reflection attack directly challenge this assumption, revealing that logical slicing alone is insufficient to prevent cross-slicing protocol tunneling if underlying resources are shared.

\subsection{\textbf{Attack Success Rate across Isolation Levels:}}
To execute the attack, the malicious \gls{ue} requires only the valid \gls{teid}-\gls{ip} pair of the victim \gls{ue} and the \gls{ip} address of any \gls{upf} serving the victim. Our experiments demonstrate that the spoofed destination \gls{ip} in the forged \gls{gtp} header can target any \gls{upf} along the \gls{pdu} path of the victim \gls{ue}, whether it is the $I-\gls{upf}$, $C-\gls{upf}$, or $P-\gls{upf}$. By setting the inner payload's destination to the \gls{dn} (or any public server) and the source to the victim \gls{ue}'s \gls{ip} address, the attacker can trick the \glspl{upf} into injecting the traffic into the victim's uplink tunnel, causing the \gls{dn} to reflect the traffic to the victim \gls{ue}.

Table~\ref{tab:reflect_results} details the attack success rate across the three deployed topologies. The first two columns in the table specify the attacker's slice and the target victim slice, respectively. As mentioned in Section~\ref{sec:reflection}, Slices 1 and 2 are selected as attacker slices, while Slices 3 and 4 are set as victim slices, and the attack is evaluated against all four pairs of configurations. The table also points out the shared (S) or isolated (I) physical and virtual resources between the attacker and victim slices. The results yield several critical insights regarding slice isolation.
\begin{itemize}
    \item \textbf{Susceptibility of Shared Resource (Topology 1):} In a fully shared environment, the reflection attack is universally successful across all slice boundaries. Even if the victim and attacker reside on slices with entirely different \gls{pdu} data paths, the attack succeeds as long as a \gls{upf} is shared between them. This is because the shared \gls{upf} contains active \gls{gtp} tunnel contexts for users across multiple slices and fails to verify the inner \gls{gtp} header contents; it blindly forwards the malicious, encapsulated traffic into the victim's uplink tunnel. This magnifies the impact of the attack and broadens the threat surface and attack targets.
    
    \item \textbf{The Illusion of \gls{vnf} Isolation (Topology 3):} Topology 3 isolates the \glspl{vnf} for the \gls{urllc} slices but deploys them on a shared physical node. Table~\ref{tab:reflect_results} shows that the attack remains completely successful against the supposedly secure and isolated \gls{urllc} slices. Our analysis reveals that because the isolated \gls{upf} containers share the same underlying physical host, they also share the internal N9 bridging interface. This shared Layer-2 bridge allows the spoofed \gls{gtp} traffic to bleed across slice boundaries, proving that \gls{vnf} isolation without physical or strict network-layer isolation fails to secure the data plane.
    
    \item \textbf{Effective Mitigation via Strict Isolation (Topology 2):} The attack was only mitigated in Topology 2, where the \gls{urllc} slices were granted strict isolation at both the \gls{vnf} and physical levels. Because of this isolation, the \glspl{upf} serving the attacker lacks the active \gls{gtp} tunnel contexts for the victim \glspl{ue} with a dedicated \glspl{upf}. Consequently, the \glspl{upf} cannot inject the spoofed traffic into the victim's uplink tunnel and successfully drop the traffic.
\end{itemize}

\subsection{\textbf{Quantifying the Impact on the Victim:}}
\begin{figure}
    \centering
    \includegraphics[width=0.4\linewidth]{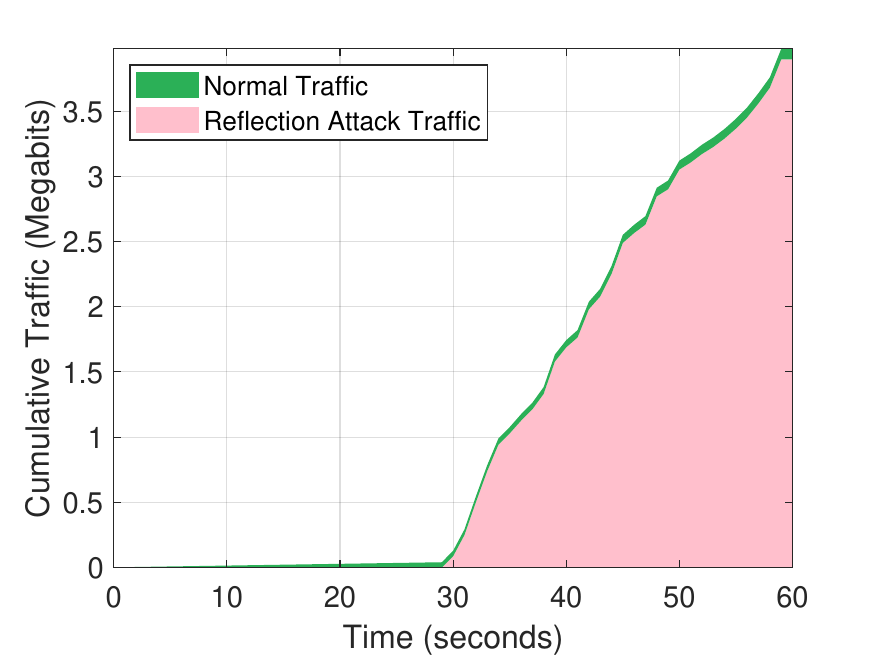}
    \caption{Network traffic going through the victim UE's tunnel. \textcolor{Green}{Green} is the normal UE traffic and \textcolor{CarnationPink}{Pink} is the reflection attack traffic injected into the Victim UE's tunnel.}
    \label{fig:injected_traffic}
\end{figure}

Figure~\ref{fig:injected_traffic} illustrates the severe volumetric impact of a successful reflection attack on a victim \gls{ue}'s tunnel. The figure shows the cumulative traffic passing through the victim \gls{ue}'s tunnel, including both the legitimate traffic from the victim \gls{ue} and the reflected attack traffic from the attacker. During normal operation (0 to 30 seconds), the legitimate traffic behaves normally. However, once the reflection attack is initiated at 30 seconds, the cumulative traffic injected into the victim's tunnel spikes violently.

Because the attacker utilizes a small spoofed query to trigger a disproportionately large response from the \gls{dn}, this amplification requires minimal effort from the attacker \cite{shaik2025uncovering}. The consequences for the victim \gls{ue} are severe: the unallocated traffic rapidly forces complete data quota depletion and results in significant charging and billing fraud, as the operator's core network attributes all the reflected data consumption directly to the victim's account.

\begin{table*}[t!]
    \centering
\caption{Results of Reflection Attack on Different Topologies and Different Isolation Levels}
\label{tab:reflect_results}
    \begin{tabular}{l|l||p{2cm}|p{2cm}||c}\toprule[1.5pt]
         \rowcolor{gray!20}
         \multicolumn{5}{c}{\textbf{Topology 1 (Shared Resources for all Slices)}}\\ \midrule
         \rowcolor{blue!7}
         & & \multicolumn{2}{c||}{\textbf{\textcolor{red}{S}hared / \textcolor{red}{I}solated Resources}} & \\ \cline{3-4}
         \rowcolor{blue!7} 
         \bf Attacker's Slice & \bf Victims's Slice & \bf Phy. & \bf VNF & \bf Attack (Un)Successful \\\hline
         Slice 1 - eMBB &  Slice 3 - eMBB &  S&  S&  \cellcolor{red!50}Successful\\\hline
         \rowcolor{blue!7}
         Slice 1 - eMBB &  Slice 4 - URLLC  &  S&  S&  \cellcolor{red!50}Successful\\\hline
         Slice 2 - URLLC  &  Slice 4 - URLLC &  S&  S&  \cellcolor{red!50}Successful\\\hline
         \rowcolor{blue!7}
         Slice 2 - URLLC  &  Slice 3 - eMBB &  S&  S&  \cellcolor{red!50}Successful\\\midrule
         \rowcolor{gray!20}
         \multicolumn{5}{c}{\textbf{Topology 2 (Isolated Phy. \& VNF for URLLC Slices Only)}}\\ \midrule 
         Slice 1 - eMBB &  Slice 3 - eMBB &  S&  S&  \cellcolor{red!50}Successful\\\hline
         \rowcolor{blue!7}
         Slice 1 - eMBB &  Slice 4 - URLLC  &  I&  I&  \cellcolor{green!50}Unsuccessful\\\hline
         Slice 2 - URLLC  &  Slice 4 - URLLC &  I&  I&  \cellcolor{green!50}Unsuccessful\\\hline
         \rowcolor{blue!7}
         Slice 2 - URLLC  &  Slice 3 - eMBB &  I&  I&  \cellcolor{green!50}Unsuccessful\\ \midrule
         \rowcolor{gray!20}
         \multicolumn{5}{c}{\textbf{Topology 3 (Isolated VNF for URLLC Slices Only)}}\\ \midrule
         Slice 1 - eMBB &  Slice 3 - eMBB &  S&  S&  \cellcolor{red!50}Successful\\\hline
         \rowcolor{blue!7}
         Slice 1 - eMBB &  Slice 4 - URLLC  &  S&  I&  \cellcolor{red!50}Successful\\\hline
         Slice 2 - URLLC  &  Slice 4 - URLLC &  S&  I&  \cellcolor{red!50}Successful\\\hline
         \rowcolor{blue!7}
         Slice 2 - URLLC  &  Slice 3 - eMBB &  S&  I&  \cellcolor{red!50}Successful\\ \hline       
         \bottomrule[1.25pt]
    \end{tabular}
\end{table*}

\begin{tcolorbox}[
    colback=blue!7,   
    colframe=blue!50!black,  
    boxrule=0.5pt,           
    arc=3pt,                 
    left=6pt, right=6pt, top=6pt, bottom=6pt 
]
\textbf{Key Takeaway:} Reflection attacks expose the illusion of software-defined security; spoofed traffic easily breaches logical slice boundaries by bleeding across shared physical interfaces (Topologies 1 and 3). Mitigating this cross-slice injection, which inflicts severe quota depletion and billing fraud, fundamentally requires strictly dedicated physical infrastructure (Topology 2).
\end{tcolorbox}

\section{Insights from Flooding Attack Experiments} \label{sec:results_flood}
To comprehensively evaluate the resilience of network slices under a flooding attack, the testbed was subjected to a continuous 10,000 \gls{pps} flooding attack. The following subsections analyze the cascading effects of this resource exhaustion across the three implemented topologies.

\subsection{Latency Degradation}
\begin{figure}[t]
    \centering
    \includegraphics[width=\linewidth]{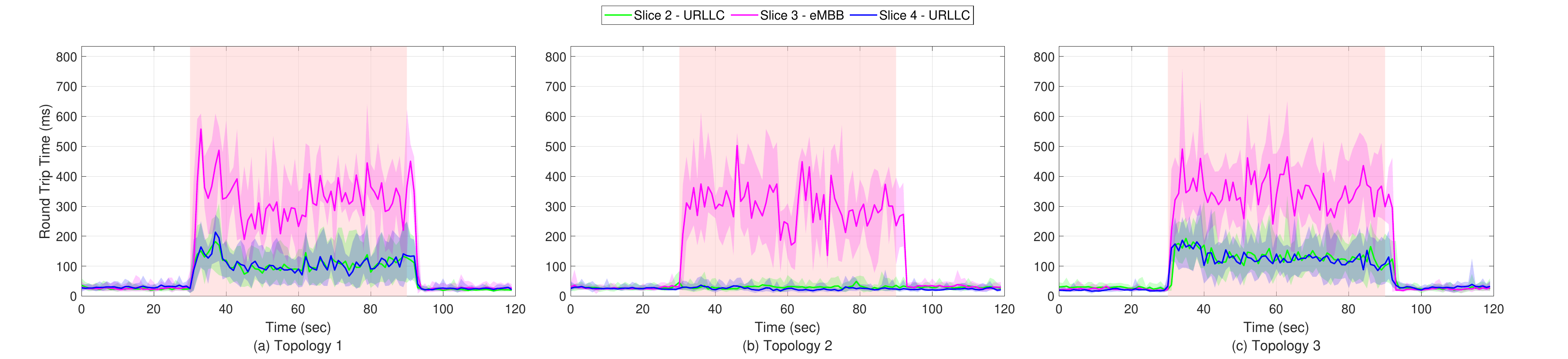}
    \caption{Round Trip Time (RTT) Results}
    \label{fig:rtt}
\end{figure}

To evaluate the impact of a flooding attack on adjacent slice latency, we monitor the \gls{rtt} of the traffic sent by the legitimate \glspl{ue} across all three testbed topologies. The results are illustrated in Figure~\ref{fig:rtt}. At the start of the experiment, we measure the \gls{rtt} of legitimate \glspl{ue} in Slices 2, 3, and 4. After $30$ seconds, the attack traffic is launched from Slice 1, which lasts for $60$ seconds and ends at $90$ seconds into the experiment. This attack window is represented in the red shaded area in Figure~\ref{fig:rtt}. Once the attack traffic stops at $90$ seconds, the experiment runs for an additional $30$ seconds with legitimate traffic to analyze any after-effects of the flooding attack. Additionally, the figure contains the average \gls{rtt} represented by solid lines, and the maximum and minimum \gls{rtt} values within a single second are shown using a shaded area around the solid line.

\textbf{Impact of Fully Shared Resources (Topology 1):} In the first experimental topology, where all slices share the same physical node and \gls{upf} instances, the baseline \gls{rtt} for all slices during normal operation ($0$ to $30$ seconds) stays steadily around $30~ms$, as depicted in Figure~\ref{fig:rtt} (a). This higher baseline delay is an inherent artifact of the software-based, user-space virtualization employed in our testbed, a limitation we formally acknowledge and detail in Section~\ref{sec:limitation}. When the flooding attack starts at the 30-second mark, the impact of the attack is immediate, resulting in catastrophic degradation of latency. Slice 3, provisioned as an \gls{embb} slice, experiences the most severe impact, with \gls{rtt} spiking drastically to peak values nearing $600~ms$. On the other hand, Slices 2 and 4 are completely unprotected in this shared configuration, and they exhibit a similar latency trend, peaking at $200~ms$ and staying around $100~ms$ throughout the attack duration. This indicates that a volumetric flooding attack targeting one slice completely exhausts the shared \gls{upf}'s computational resources and interface queues, cascading the delay degradations across all active legitimate \gls{ue} sessions across slices (cross-slice impact). The severity of the \gls{rtt} degradation depends heavily on the shared data path. Slice 3 experiences the highest delay because it shares the exact same data plane path to the \gls{dn} as the attacking slice. Slices 2 and 4, however, only share the $I-\gls{upf}$ with the attacker, resulting in a distinctly lower latency impact.

\textbf{Effectiveness of Strict Physical and \gls{vnf} Isolation (Topology 2):} Figure~\ref{fig:rtt} (b) illustrates the network's behavior when strict isolation is requested by \gls{urllc} slices at both the \gls{vnf} and physical node levels. The baseline \gls{rtt} in Topology 2 is similar to Topology 1. During the attack, the \gls{ue} in Slice 3 continues to suffer severe latency degradation, averaging around $300~ms$, because it shares the core infrastructure with Slice 1, which the attacking \gls{ue} is targeting. However, Slices 2 and 4 remain entirely unaffected during the flooding attack. Their \gls{rtt} maintain a stable baseline of approximately $30~ms$ throughout the entire 60-second attack duration. This demonstrates that decoupling the data plane \glspl{upf} at both the software container and physical hardware layers effectively contains volumetric congestion, preserving the deterministic latency required for critical communications.

\textbf{The Susceptibility of Partial Isolation (Topology 3):} The most revealing insights regarding architectural bottlenecks emerge from Topology 3, where Slices 2 and 4 are allocated dedicated \gls{upf} containers (\gls{vnf} isolation) but remain hosted on the same physical node as the targeted Slice 1. As shown in Figure~\ref{fig:rtt} (c), while Slice 3 predictably spikes to roughly $600~ms$, Slice 2 and Slice 4 fail to maintain their baseline latency. Instead, their \gls{rtt} metrics degrade significantly, closely mirroring the approximate $200~ms$ spikes observed in the fully shared Topology 1. This empirical evidence proves that logical \gls{vnf} isolation is entirely negated if the underlying physical hypervisor, specifically shared CPU cycles, memory bandwidth, and virtual Layer-2 bridges, becomes saturated by a volumetric flood attack.

\textbf{Implications for Next-Generation SLAs:} To contextualize the severity of the $\sim200~ms$ latency spikes observed in \gls{urllc} slices in Topologies 1 and 3, it is critical to evaluate these metrics against standardized \glspl{sla}. According to \gls{3gpp} TS 22.261, \gls{urllc} applications mandate strict end-to-end latency budgets typically ranging from $1~ms$ to $10~ms$ \cite{3gpp.22.261}. While our software-based testbed environment establishes a higher baseline latency ($\sim30 ms$) as noted previously, the critical insight lies in the relative degradation: the volumetric flood caused a severe $400\%$ to $600\%$ increase in latency across the shared slices. In a commercial, hardware-accelerated \gls{5g} deployment operating at a $1~ms$ baseline, this proportional loss of determinism would result in a catastrophic \gls{sla} violation, causing automated emergency braking failures in \gls{v2x} networks or triggering safety halts in Industry 4.0 cooperative robotics. This starkly highlights that without the optimization-driven strict physical isolation demonstrated in Topology 2, shared data planes fundamentally lack the structural resilience required to safely host mission-critical slices.

\textbf{The Role of Jitter and Determinism in network Slicing:} In \gls{5g} and next-generation mobile architectures, maintaining strict isolation between network slices is paramount for preserving \gls{qos}. While average latency is a standard performance metric, bounding jitter is often more critical than minimizing average \gls{rtt} for \gls{urllc} slices \cite{3gpp.22.104}. \gls{urllc} traffic relies on strict determinism, requiring the network to guarantee packet delivery within a tightly bounded, predictable time window. To evaluate the resilience of determinism under varying network conditions, we analyzed the delay variations across the three topologies.

\begin{figure}[t]
    \centering
    \includegraphics[width=\linewidth]{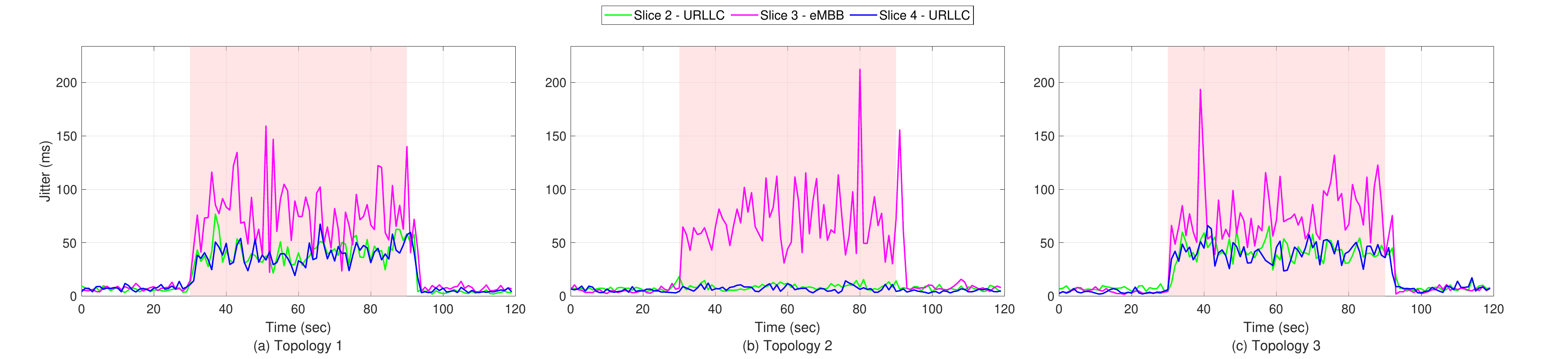}
    \caption{Jitter Results}
    \label{fig:jitter}
\end{figure}

\begin{itemize}
    \item \textbf{Topology 1 (Fully Shared Resources):} Figure~\ref{fig:jitter} (a) illustrates the jitter results in Topology 1, where all slices share the physical and virtual resources. The baseline jitter before the attack starts is stable for all slices. When the attack starts at $30$ seconds, a severe degradation in jitter is evident for all slices. Slice 3 experiences extreme fluctuations, frequently spiking between $100~ms$ and $150~ms$. Slices 2 and 4 are also heavily impacted by the shared congestion, with their jitter oscillating unpredictably between $25~ms$ and $75~ms$ during the attack. This demonstrates a complete loss of the determinism required for \gls{urllc} traffic when data plane resources are fully shared.
    
    \item \textbf{Topology 2 (Strict Physical and VNF Isolation):} With strict physical and \gls{vnf} level isolation of \gls{urllc} slices (Slices 2 and 4), the jitter remains stable even throughout the attack window, as shown in Figure~\ref{fig:jitter} (b). On the other hand, Slice 3 shows a similar fluctuating trend to Topology 1 results. This proves that decoupling the data plane at both hardware and \gls{vnf} levels successfully shields critical deterministic routing from volumetric floods.
    
    \item \textbf{Topology 3 (VNF Isolation Only):} In this hybrid configuration, the failure of partial isolation is highly visible. While Slice 3 spikes aggressively, the variance in packet delivery for Slices 2 and 4 also degrades significantly, as shown in Figure~\ref{fig:jitter} (c). Despite having isolated and dedicated \gls{upf} containers, these slices experience persistent jitter fluctuating around $50~ms$ during the attack. This further proves that \gls{vnf} isolation alone is insufficient, as underlying host resource contention causes \gls{urllc} traffic to lose the strict determinism required for critical applications.
\end{itemize}

This loss of determinism and high variance in packet delivery would cause critical industrial automation, remote surgery, or autonomous vehicle applications to fail, as these use cases depend on tightly synchronized time-sensitive control loops \cite{3gpp.22.104, ITU2083}.

\begin{tcolorbox}[
    colback=blue!7,   
    colframe=blue!50!black,  
    boxrule=0.5pt,           
    arc=3pt,                 
    left=6pt, right=6pt, top=6pt, bottom=6pt 
]
\textbf{Key Takeaway:} Volumetric flooding proves that software-only \gls{vnf} isolation (Topology 3) cannot protect critical \gls{urllc} traffic if physical resources are shared. The resulting latency spikes (up to 600\%) and severe jitter destroy deterministic \glspl{sla}, risking catastrophic cyber-physical failures. Only strict physical hardware isolation (Topology 2) guarantees mission-critical determinism.
\end{tcolorbox}

\subsection{Service Reliability (Packet Loss)}
\begin{figure}[t]
    \centering
    \includegraphics[width=0.5\linewidth]{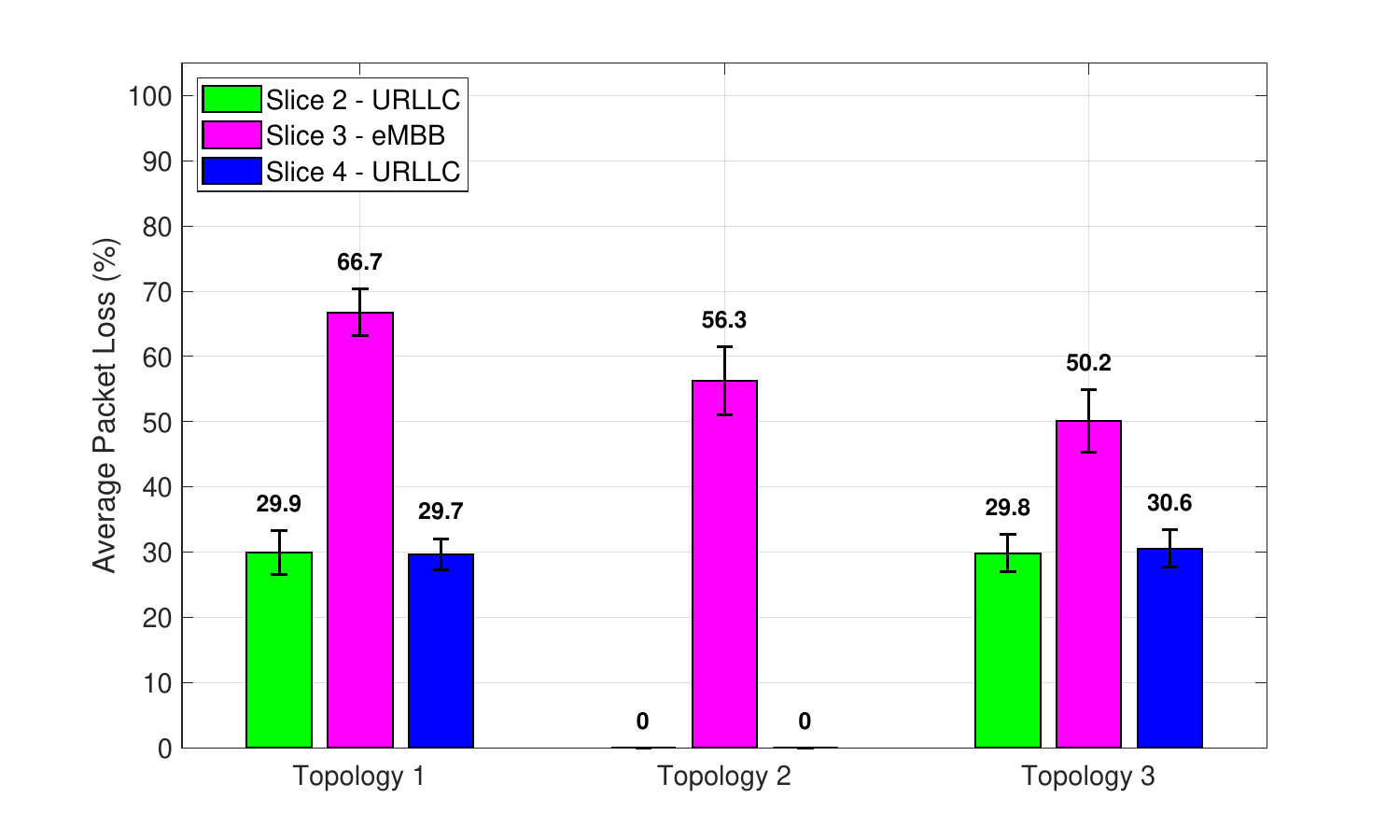}
    \caption{Average Packet Loss Results}
    \label{fig:loss}
\end{figure}

While latency and jitter quantify the degradation of active communication, measuring average packet loss directly evaluates the network's ability to maintain service reliability and availability during volumetric floods. Figure~\ref{fig:loss} illustrates the average packet loss percentages along with $95\%$ confidence intervals across three testbed topologies under the 10,000 \gls{pps} flood attack.

\textbf{Susceptibility of Fully Shared Infrastructure (Topology 1):} In the fully shared configuration, the flood attack severely compromises service reliability across all network slices. Slice 3 sustained the most catastrophic impact, experiencing an average packet loss of $66.7\%$. This extreme loss is justified by the routing path, because Slice 3 shares the exact same data plane path to the \gls{dn} as the attacking slice; its traffic competes directly within the most congested queues of the shared \glspl{upf}. However, the \gls{urllc} slices are also severely compromised, with Slice 2 and Slice 4 exhibiting $29.9\%$ and $29.7\%$ packet loss, respectively. Because all slices rely on a single, shared $I-\gls{upf}$ instance, the flood exhausts the \gls{upf} container’s processing capacity and fills its interface buffers, forcing the \gls{upf}  to indiscriminately drop incoming packets regardless of their slice assignment.

\textbf{Complete Protection via Strict Isolation (Topology 2):} The empirical data from Topology 2 provides the strongest validation for the necessity of strict physical and \gls{vnf} isolation. While Slice 3 continues to suffer severe degradation with an average packet loss of $56.3\%$ due to its shared path with the attacking \gls{ue}, the \gls{urllc} slices are perfectly insulated. Both Slice 2 and Slice 4 experience $0\%$ packet loss during the attack duration. By dedicating separate \gls{upf} containers hosted on entirely distinct physical nodes, the network effectively decouples the processing queues and hardware resources. The targeted flood exhausts the resources of the \gls{embb} \glspl{upf} but cannot traverse the physical boundary to starve the \gls{urllc} \glspl{upf}, guaranteeing absolute reliability for critical traffic.

\textbf{The Failure of VNF-Only Isolation (Topology 3):} The results from Topology 3 expose a critical architectural weakness in deploying \gls{5g-advanced} and beyond network slices purely through software-defined boundaries. Although Slices 2 and 4 are provisioned with dedicated \gls{upf} instances (\gls{vnf} isolation), they are hosted on the same physical server as the attacker's Slice 1. The result reveals that Slice 2 and Slice 4 suffer $29.8\%$ and $30.6\%$ average packet loss, respectively. Statistically, this is virtually identical to the $29.9\%$ and $29.7\%$ loss observed in the completely unisolated Topology 1. Meanwhile, Slice 3 experiences an average loss of $50.2\%$. This proves that logical \gls{vnf} isolation is fundamentally ineffective against volumetric data plane flood attacks. When the shared physical hypervisor, including host CPU cycles, memory bandwidth, and virtual network interface controllers (NICs), becomes saturated by the flood, the host operating system begins dropping packets before they can even be routed to the isolated \gls{urllc} slice \gls{upf} containers.  

\textbf{Implications for Next-Generation SLAs:} The operational impact of the observed packet loss extends far beyond simple bandwidth reduction. The \gls{3gpp} standardized reliability target for \gls{urllc} slices is aggressively defined as $99.999\%$ successful packet delivery \cite{3gpp.22.104}. While our optimization model focuses on resource allocation with security requirements rather than explicit \gls{qos} traffic shaping, it provisions the architectural isolation required to prevent cross-slice resource contention. The empirical results demonstrate that Slices 2 and 4 sustained approximately $30\%$ packet loss in both the fully shared configuration (Topology 1) and the \gls{vnf}-isolated configuration (Topology 3). This represents a total collapse of the theoretical $99.999\%$ reliability \gls{sla}, proving that a volumetric data plane attack on a co-allocated \gls{embb} slice can easily trigger a \gls{dos} state for adjacent, high-priority slices. Consequently, achieving and safeguarding the standardized reliability targets for \gls{5g-advanced} and beyond absolutely necessitates the proactive, physically isolated resource allocation strategies proposed by our model.

\begin{tcolorbox}[
    colback=blue!7,   
    colframe=blue!50!black,  
    boxrule=0.5pt,           
    arc=3pt,                 
    left=6pt, right=6pt, top=6pt, bottom=6pt 
]
\textbf{Key Takeaway:} \gls{vnf} isolation on shared hardware (Topology 3) completely fails to protect service reliability during volumetric flooding, resulting in $\sim30\%$ packet loss for mission-critical slices. This massive data drop shatters the $99.999\%$ \gls{urllc} reliability standard. Consequently, guaranteeing next-generation \glspl{sla} strictly requires the dedicated physical isolation demonstrated in Topology 2.
\end{tcolorbox}

\subsection{Bandwidth Starvation (TCP Throughput)}
\begin{figure}
    \centering
    \includegraphics[width=\linewidth]{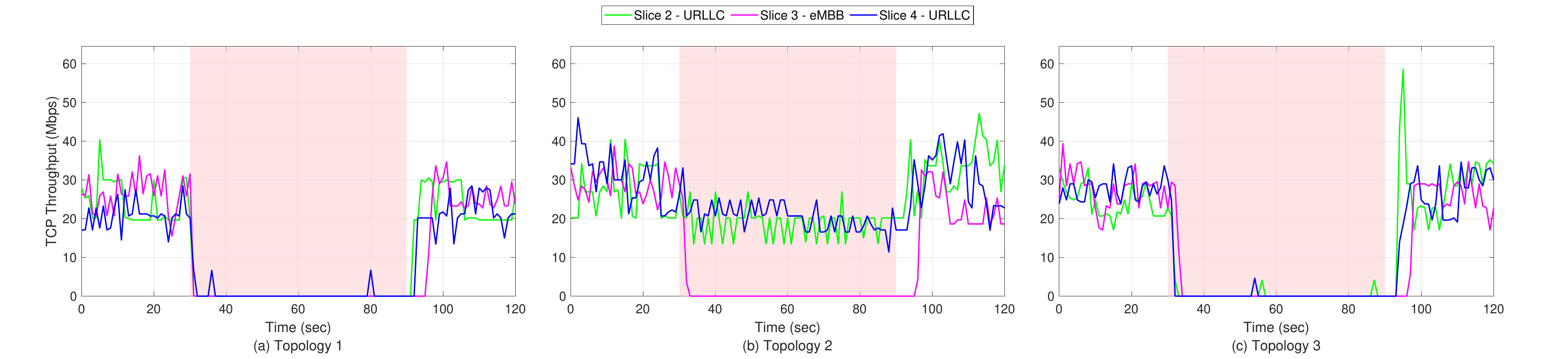}
    \caption{Throughput Results}
    \label{fig:Throughput}
\end{figure}

While \gls{udp} probing effectively measures network-layer latency and packet loss, measuring \gls{tcp} throughput quantifies usable application-layer bandwidth. Because \gls{tcp} utilizes a closed-loop congestion control, it is highly sensitive to the underlying network degradation observed in the previous results. Figure~\ref{fig:Throughput} illustrates the \gls{tcp} throughput for all three slices before, during, and after the flooding attack traffic.

\textbf{Total Collapse in Shared Infrastructure (Topology 1):} Prior to the attack, all slices maintain a fluctuating steady-state baseline approximately between $15~Mbps$ and $40~Mbps$, as shown in Figure~\ref{fig:Throughput} (a). However, upon the initiation of the flooding attack at the 30-second mark, the \gls{tcp} throughput for all three slices instantly collapses to $0~Mbps$. Transport-layer telemetry reveals the underlying mechanics of this starvation: the initial wave of packet loss triggers an immediate surge in \gls{tcp} retransmissions (surpassing $150$ retransmissions per second). As consecutive retransmissions continue to drop at the overwhelmed $I-\gls{upf}$ queues, the \gls{tcp} sender transitions into exponential \gls{rto} backoff, collapsing the \gls{cwnd} from an operating baseline of $\sim100 - 300~KB$ down to minimum segment size. Notably, Figure~\ref{fig:Throughput} (a) contains minor throughput spikes observed for Slice 4 during the attack window (e.g., at $\sim36s$ and $\sim80s$). This indicates that some \gls{tcp} traffic may survive the congestion caused by the flooding attack before the connection backs off again due to subsequent packet drops.

\textbf{Guaranteed Availability via Strict Isolation (Topology 2):} Topology 2 demonstrates the efficacy of physical and \gls{vnf} isolation. While Slice 3 (\gls{embb}) suffers complete starvation due to sharing the attacked data path, the isolated \gls{urllc} slices (Slice 2 and 4) maintain active data transmission during the attack phase. Although experiencing a minor throughput reduction to the lower bound of their $15 -  30~Mbps$ baseline, both \gls{urllc} slices preserve uninterrupted data plane availability. In contrast to the unisolated configurations, analyzing the \gls{cwnd} and retransmissions reveals that Slices 2 and 4 sustain an uncollapsed \gls{cwnd} (ranging between $100~KB$ and $300~KB$) and register negligible retransmission activity throughout the attack. 

\textbf{The Illusion of VNF Isolation and TCP Bursting (Topology 3):} In Topology 3, despite Slices 2 and 4 utilizing dedicated \gls{upf} containers, the saturation of the shared underlying physical host CPU and virtual switches causes their \gls{tcp} throughput to plummet to $0~Mbps$ and trigger a \gls{cwnd} collapse, identically mirroring the failure of Topology 1. However, Figure~\ref{fig:Throughput} (c) reveals a distinct post-attack phenomenon: immediately following the attack at the $90$-second mark, Slice 2 experiences a sharp throughput burst reaching nearly $60~Mbps$ before returning to baseline throughput. This transient overshoot occurs because the sender's transmission queue, heavily backlogged during the $60$-second starvation window, is flushed simultaneously as the physical host queues clear, driving the \gls{cwnd} through an aggressive slow-start recovery phase.

\textbf{Implications for Next-Generation SLAs:} Complete bandwidth starvation in Topologies 1 and 3 illustrates the operational risks for critical services in \gls{5g-advanced} and Beyond. For mission-critical use cases, such as real-time video telemetry or industrial control loops, a drop to $0~Mbps$ constitutes a complete service outage. These empirical findings demonstrate the necessity of a ``Slice Isolation-as-a-Service'' model: because logical slicing over shared hardware cannot prevent transport-layer collapse during data plane saturation, high-criticality tenants must be provisioned with dedicated physical resources to guarantee SLA continuity under adversarial conditions.

\begin{tcolorbox}[
    colback=blue!7,   
    colframe=blue!50!black,  
    boxrule=0.5pt,           
    arc=3pt,                 
    left=6pt, right=6pt, top=6pt, bottom=6pt 
]
\textbf{Key Takeaway:} Shared hardware environments (Topologies 1 and 3) completely fail to sustain application-layer throughput during volumetric floods, as overwhelmed queues trigger severe \gls{rto} backoff and \gls{cwnd} collapse. Preventing this $0~Mbps$ service outage for mission-critical slices strictly requires the dedicated physical isolation demonstrated in Topology 2.
\end{tcolorbox}

\section{Performance Evaluation of the Optimization Model} \label{sec:perf_eval}
To comprehensively evaluate the scalability and trade-offs of the proposed resource allocation model, the \gls{minlp} formulation was implemented in Julia \cite{bezanson2017julia} using the JuMP modeling language \cite{dunning2017jump} and solved utilizing the \gls{scip} optimization solver \cite{scip_web, scip_git}. All computational experiments were executed on a Linux-based environment equipped with a 32-core Intel processor and 32 GB of RAM. The evaluation compares the performance of a standard ``Non-Secure Allocation,'' a baseline scenario where all network slices are assigned a low security tier, effectively bypassing the isolation constraints, against the proposed ``Secure Allocation,'' which actively enforces the Slice Isolation-as-a-Service security constraints to mathematically guarantee the physical and virtual separation required by high-priority slices.

\begin{figure}[t]
\centering
\begin{tikzpicture}[
    font=\footnotesize, 
    startstop/.style={rectangle, rounded corners, minimum width=3cm, minimum height=0.6cm, text centered, draw=black, fill=gray!10, align=center},
    process/.style={rectangle, minimum width=3.5cm, minimum height=0.8cm, text centered, draw=black, fill=blue!5, align=center},
    decision/.style={diamond, aspect=2, minimum width=2.5cm, minimum height=0.5cm, text centered, draw=black, fill=orange!10, align=center},
    arrow/.style={thick, ->, >=stealth, align=center}
]

\node (start) [startstop] {Initialize Optimization Model};
\node (phase1) [process, below=0.4cm of start] {\textbf{Phase 1: Strict Optimality}\\ Gap = 0.0\% $|$ Time Limit = 1800s};
\node (opt1) [process, below=0.4cm of phase1] {Execute Solver (\texttt{optimize!})};
\node (check) [decision, below=0.4cm of opt1] {Status == \\ \texttt{OPTIMAL}?};
\node (phase2) [process, right=2.5cm of check] {\textbf{Phase 2: Relaxed Gap}\\ Rebuild Model \\ Gap = 1.0\% $|$ Time Limit = $\infty$};
\node (opt2) [process, below=0.4cm of phase2] {Execute Solver (\texttt{optimize!})};
\node (end) [startstop, below=1.0cm of check] {Output Results \\ \& Terminate};

\draw [arrow] (start) -- (phase1);
\draw [arrow] (phase1) -- (opt1);
\draw [arrow] (opt1) -- (check);
\draw [arrow] (check) -- node[anchor=east] {Yes} (end);
\draw [arrow] (check) -- node[anchor=south, yshift=0.1cm] {No\\(Time Limit Reached)} (phase2);
\draw [arrow] (phase2) -- (opt2);
\draw [arrow] (opt2) |- (end);

\end{tikzpicture}
\caption{Two-phase solver execution logic balancing absolute mathematical optimality against computational time constraints.}
\label{fig:solver_logic}
\end{figure}
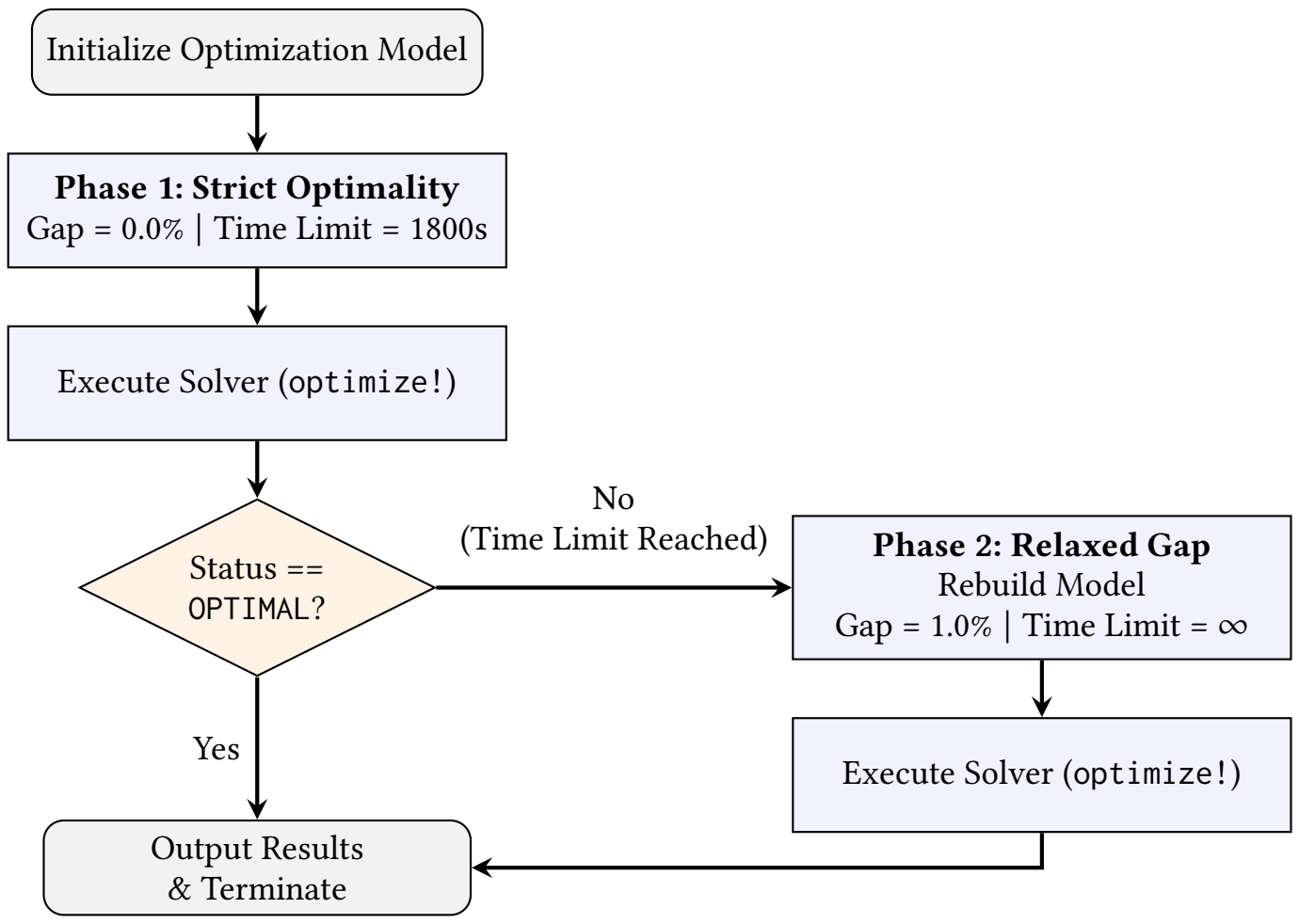

Because \gls{minlp} problems scale exponentially in complexity as physical nodes and network slices increase, relying on absolute mathematical optimality (a $0\%$ optimality gap) can lead to very high execution times. To balance computational feasibility with resource efficiency, the optimization model implements a two-phase execution strategy, as illustrated in Figure~\ref{fig:solver_logic}. In Phase 1, the solver targets strict optimality by setting a $0\%$ gap tolerance while enforcing a strict maximum execution time of $1800$ seconds ($30$ minutes). If the model complexity allows the solver to find the absolute optimal solution within this time window, the execution stops successfully. However, if the solver reaches the $1800$-second limit before proving absolute optimality, it terminates and initiates Phase 2. To avoid residual state errors in the JuMP environment, the model is entirely re-initialized in this phase. In this secondary phase, the time limit is effectively removed (set to infinity), and the optimality gap is relaxed to $1\%~(0.01)$. This strategic relaxation ensures that the optimization model eventually converges on a highly optimal resource allocation without hanging indefinitely on fractions-of-a-percent improvements.

\subsection{Computational Complexity and the Time Trade-Off}
\begin{figure}[t]
    \centering
    \begin{subfigure}{0.24\textwidth}
        \centering
        \includegraphics[width=\linewidth]{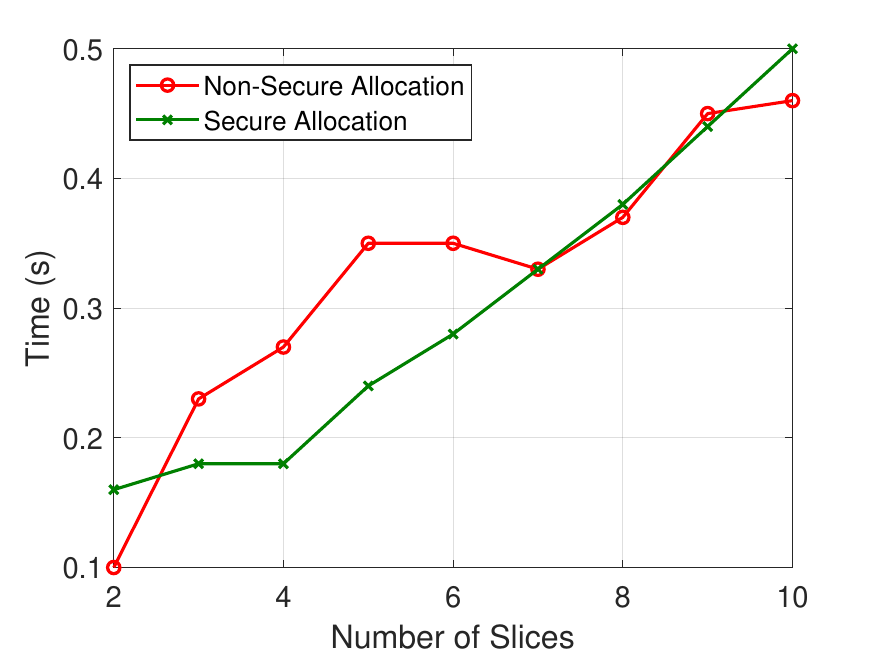}
        \caption{Model Presolve Time}
        \label{fig:perf_presolve}
    \end{subfigure}
    \begin{subfigure}{0.24\textwidth}
        \centering
        \includegraphics[width=\linewidth]{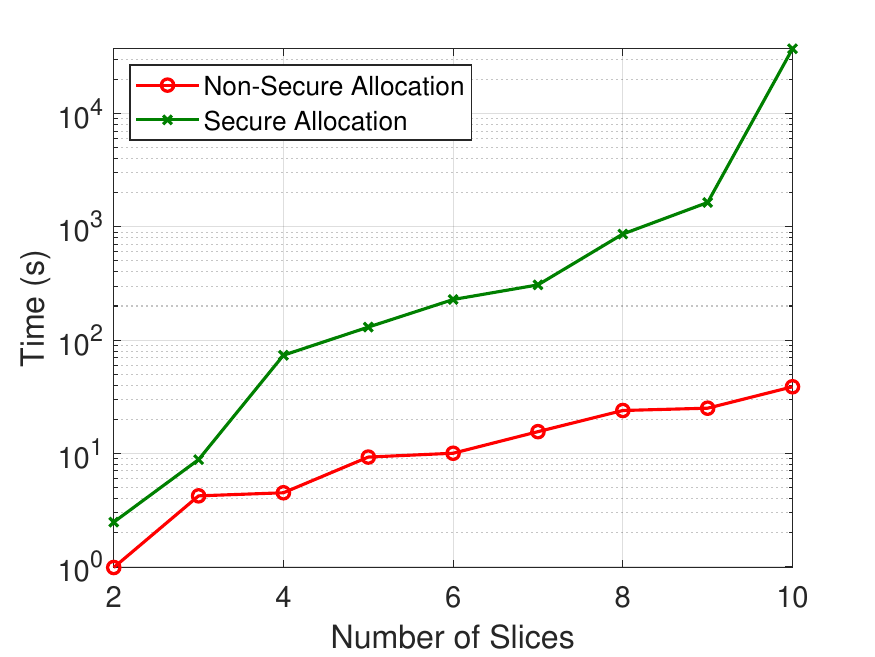}
        \caption{Model Solve Time}
        \label{fig:perf_solve}
    \end{subfigure}
    \begin{subfigure}{0.24\textwidth}
        \centering
        \includegraphics[width=\linewidth]{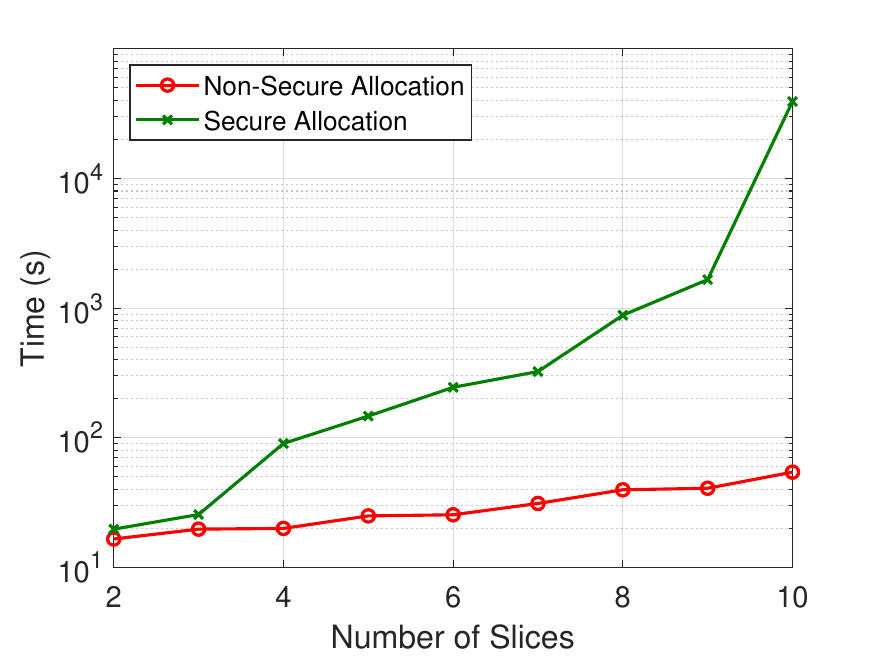}
        \caption{Total Model Run Time}
        \label{fig:perf_run_time}
    \end{subfigure}
    \begin{subfigure}{0.24\textwidth}
        \centering
        \includegraphics[width=\linewidth]{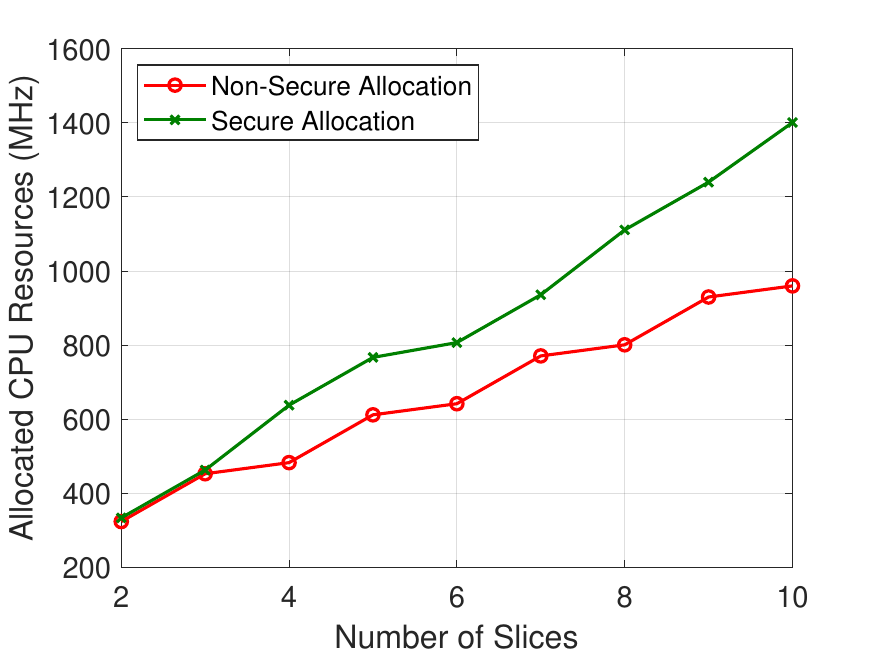}
        \caption{Allocated CPU Resources}
        \label{fig:cpu}
    \end{subfigure}
    
    \caption{Performance Evaluation of the Optimization Model}
    \label{fig:perf_opt_model}
\end{figure}

When evaluating the scalability of the optimization model, understanding the computational overhead introduced by security constraints is critical. Table \ref{tab:sim_paramter}(b) lists the parameters and their corresponding values used in evaluating the performance of the optimization model. All parameter values from Table~\ref{tab:sim_paramter}(a) were used except the number of physical nodes and network slices, which were set to $5$ and varied from $2$ to $10$, respectively. Figures \ref{fig:perf_presolve}, \ref{fig:perf_solve}, and \ref{fig:perf_run_time} collectively illustrate the time-based metrics, including presolve time, model solve time, and total run time, as the network scales from $2$ to $10$ slices.

Figure~\ref{fig:perf_presolve} demonstrates that the model presolve time scales relatively linearly and remains trivial for both approaches, peaking at merely $0.5$ seconds even for $10$ slices. This indicates that the initial generation of the mathematical model, including variable bounding and basic assignment of constraints within JuMP, is highly efficient and does not create a computational bottleneck.

However, the true computational cost of strict isolation becomes glaringly apparent during the model execution phase. Figures~\ref{fig:perf_solve} and \ref{fig:perf_run_time} reveal a massive divergence in the \gls{scip} solver's performance. The non-secure allocation scales efficiently, completing total model solve times in under $10^2$ seconds even for $10$ slices. In stark contrast, the secure allocation triggers an exponential explosion in model solve time. As observed on the logarithmic y-axis, calculating the optimal allocation for $10$ secure slices demands over $10^4$ seconds (nearly $3$ hours) of total run time.

This extreme disparity highlights the primary mathematical trade-off of proactive security. Enforcing strict physical isolation requires the optimization model to navigate a highly complex combinatorial space. The binary assignment constraints must rigorously prevent \gls{vnf} co-location across shared physical nodes, turning a standard resource allocation problem into a computationally heavy, tightly constrained optimization problem. Despite this massive increase in model solving and run times, this trade-off is justified by the absolute security it provides against the cross-slice data plane attacks demonstrated in Section~\ref{sec:results_gtp} and \ref{sec:results_flood}.

\subsection{The Resource Cost of Isolation}
In addition to the extended execution time required to solve the optimization model, enforcing strict security constraints demands a significantly larger physical hardware footprint. Figure~\ref{fig:cpu} compares the total allocated CPU resources (in MHz) in both non-secure and secure allocations as the network scales from 2 to 10 slices.

For a small network of 2 slices, the CPU requirement is nearly identical for both approaches, approximately $300~MHz$. However, as the number of slices increases, the resource gap widens considerably. At 10 slices, the non-secure allocation requires under $1000~MHz$, while the secure allocation consumes approximately $1400~MHz$, an increase of nearly $50\%$. This significant overhead occurs because the secure allocation mathematically prevents \glspl{vnf} of high-priority slices from being hosted on shared physical nodes. By eliminating the ability to pool resources and share idle CPU cycles across multiple slices, the system is forced to provision dedicated, isolated hardware, which inherently drives up total CPU consumption.

This combined overhead---both the extended time required to solve the complex resource allocation problem and the extra physical CPU resources demanded by dedicated hardware---forms the economic basis for the proposed ``Slice Isolation-as-a-Service'' model. Because providing secure allocations consumes significantly more system processing time and physical server capacity, network operators have a clear justification to assign this overhead a monetary value. By passing these operational costs (\gls{opex}) onto the tenants, operators can charge a premium to high-criticality users (such as military or medical facilities) who mandate mathematically guaranteed physical isolation, ensuring the network remains both secure and financially sustainable.

\section{Limitations of Current Work}\label{sec:limitation}
While this study successfully demonstrates the mathematical and empirical necessity of proactive network slice isolation, we acknowledge several limitations in the current implementation that pave the way for future research.

\subsection{Offline Optimization and Manual Deployment}
The proposed resource allocation optimization model currently operates as an offline, static optimization tool rather than a real-time, closed-loop orchestrator. Consequently, the secure resource allocations and architectural topologies generated by the mathematical model are currently manually configured and deployed within the \gls{5g} testbed environment. Automating the translation from mathematical output to dynamic network deployment remains a significant limitation of the current work. Future work could leverage advanced cloud-native orchestration frameworks, such as the Linux Foundation's Project Nephio \cite{nephio}, to automatically deploy these optimized, intent-driven topologies directly into Kubernetes-based \gls{5g} infrastructures. 

\subsection{Testbed Environmental Constraints and Virtualization Overhead} 
As noted during the empirical evaluation in Section~\ref{sec:results_flood}, the \gls{5g} testbed exhibits a baseline \gls{rtt} of approximately $30~ms$, which exceeds the stringent $1-10~ms$ latency budgets standardized for commercial \gls{urllc} traffic. This relatively higher baseline delay is due to the virtualized deployment of all \gls{5g} components on a single physical server. Specifically, both the \gls{5g} core and \gls{ran} emulators are deployed as software-based instances rather than hardware-accelerated appliances. Because the containerized \glspl{upf} operate in user space without hardware acceleration techniques (such as the \gls{dpdk}), the system incurs unavoidable kernel-to-user space context-switching delays. However, while the absolute baseline latency is inflated by this virtualization overhead, the relative degradation observed during volumetric attacks (e.g., $400\%$ to $600\%$ latency spikes) accurately validates the cross-slice security susceptibilities and the effectiveness of the proposed slice isolation constraints.

\subsection{Abstraction of Financial and Energy Metrics}
The optimization formulation is strictly designed to evaluate the architectural trade-offs of security isolation. The objective function minimizes total computational capacity, baseline provisioning overhead, and end-to-end traffic delay; however, it abstracts away explicit financial (\gls{capex}/\gls{opex}) and energy consumption metrics. Incorporating real-time monetary costs or server power consumption constraints would drastically increase the complexity of the \gls{minlp} model without contributing to the core security analysis. Consequently, determining the exact monetary pricing structures for the varying security tiers of the Slice Isolation-as-a-Service framework is treated as a subsequent policy and business decision left to the discretion of individual network operators.

\section{Ethical Considerations} \label{sec:ethical}
All experimental evaluations and \gls{dp} attacks in this study were conducted in a localized, privately hosted \gls{5g} testbed. To guarantee absolute containment and ensure that no malicious traffic could escape the controlled environment, the \glspl{vm} hosting the \gls{5g} testbed were entirely isolated from the public internet. Furthermore, this work does not introduce any novel attack vectors or zero-day exploits. The implemented attacks, specifically the volumetric \gls{dp} flooding and the \gls{gtp}-in-\gls{gtp} reflection exploit, rely on previously published and academically acknowledged vulnerabilities. These known exploits were reproduced in our isolated environment solely to empirically validate the defensive efficacy of the proposed Slice Isolation-as-a-Service framework.

\section{Conclusion} \label{sec:conclusion}

The empirical results from our testbed experiments expose a fundamental paradox in next-generation network slicing: while maximizing shared infrastructure is highly resource-efficient, it creates an unacceptable cross-slice threat surface. Furthermore, the total failure of software-only boundaries proves that logical \gls{vnf} separation is fundamentally incapable of withstanding volumetric data plane attacks and spoofed \gls{gtp} attacks on shared physical hosts. While strict physical isolation successfully guarantees service reliability and deterministic latency for mission-critical slices, blindly provisioning dedicated hardware for every network slice is financially prohibitive and negates the multiplexing benefits of virtualization.

This architectural trade-off strictly necessitates the mathematical optimization model formulated in the first half of this paper. Rather than relying on static, human-configured resource boundaries, the proposed mixed-integer optimization model mathematically computes the safest, most resource-efficient allocation of both physical and virtual components. By formulating strict physical isolation as a quantifiable constraint, the model enables operators to selectively provision dedicated hardware exclusively for high-priority tenants, acting as the core computational engine for the proposed ``Slice Isolation-as-a-Service'' framework. Therefore, integrating proactive, optimization-driven resource allocation provides network operators with a quantifiable mechanism to monetize mission-critical isolation and safely host heterogeneous services in evolving \gls{5g-advanced} and \gls{6g} cellular networks.

\begin{acks}
This work was supported by the Natural Sciences and Engineering Research Council of Canada (NSERC). This research was enabled in part by support provided by Research Computing Services (https://carleton.ca/rcs) at Carleton University.
\end{acks}


\bibliographystyle{ACM-Reference-Format}
\bibliography{library}

\appendix
\section{Appendix} \label{sec:appendix}
\subsection{Structure of Malformed GTP-in-GTP Packet}
\begin{figure}[thb]
    \centering
    \includegraphics[width=0.5\linewidth]{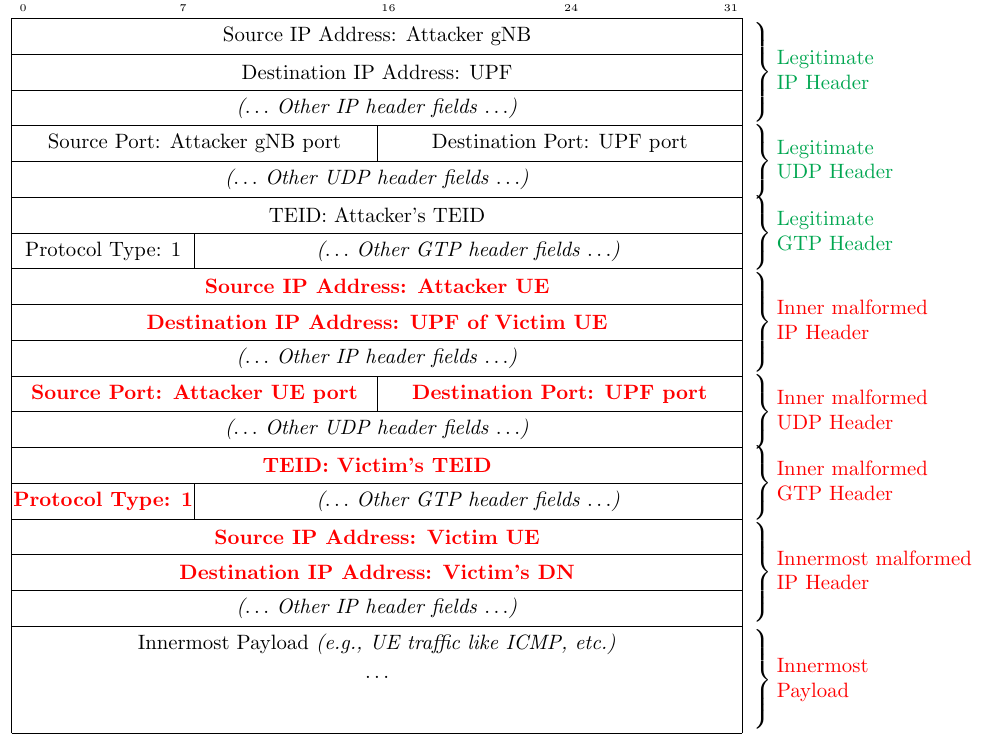}
    \caption{Structure of Malformed GTP-in-GTP Packet}
    \label{fig:GTP_packet}
\end{figure}

Figure~\ref{fig:GTP_packet} illustrates the nested protocol architecture of the malformed packet utilized to execute the cross-slice \gls{gtp}-in-\gls{gtp} attack. By exploiting the lack of robust deep packet inspection at the data plane, the attacker encapsulates a forged \gls{gtp} tunnel within a legitimately established \gls{gtp} session.

To successfully bypass standard routing boundaries and reflect malicious traffic into the victim's network slice, the attacker must precisely craft specific fields across the nested headers:  

\begin{itemize}
    \item \textbf{Outer Legitimate Headers (\gls{gnb} to \gls{upf}):} The outermost \gls{ip} and \gls{udp} headers utilize the attacker's \gls{gnb} as the source and the target \gls{upf} as the destination. The outer \gls{gtp} header contains the Attacker's \gls{teid}, ensuring the initial packet is legitimately accepted and decapsulated by the receiving \gls{upf}.
    \item \textbf{Inner Malformed Headers (The Spoof):} Nested inside the payload of the first \gls{gtp} tunnel is a second, forged set of headers. The inner \gls{ip} header spoofs the Attacker \gls{ue} as the source and targets the Victim \gls{ue}'s \gls{upf} as the destination. Crucially, the inner \gls{gtp} header contains the Victim's \gls{teid}. This tricks the \gls{upf} into mapping the subsequent payload to the victim's active session.
    \item \textbf{Innermost Payload (The Reflection):} The innermost \gls{ip} header spoofs the Victim \gls{ue} as the source \gls{ip} and targets the Victim's \gls{dn} as the destination \gls{ip}. Upon processing the inner \gls{teid}, the \gls{upf} forwards this innermost payload (e.g., an \gls{icmp} flood) directly into the victim's isolated network, effectively executing the cross-slice attack.
\end{itemize}

\end{document}